\documentclass[prd,aps,twocolumn,a4paper,showkeys,nofootinbib]{revtex4-1}

\renewcommand\section{\paragraph}

\usepackage{graphicx,psfrag}
\usepackage{mathrsfs}
\usepackage{amsmath,amsfonts,amssymb}
\usepackage{multirow}
\usepackage{comment}
\usepackage{ulem}
\usepackage{hyperref}
\usepackage{enumitem}
\usepackage{tcolorbox}

\newcommand{\be}{\begin{equation}}
\newcommand{\ee}{\end{equation}}
\newcommand{\bea}{\begin{eqnarray}}
\newcommand{\eea}{\end{eqnarray}}
\newcommand{\bel}{\begin{align}}
\newcommand{\eel}{\end{align}}

\newcommand{\orcid}[1]{\href{https://orcid.org/#1}{
\includegraphics[width=10pt]{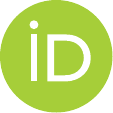}
}}

\def\Msun{{\rm M_{\odot}}}
\def\Mo{{\Msun}}

\def\GMc2{{\rm G M_{\odot} c^{-2}}}

\def\Egw{E_\text{rad}}
\def\Jgw{J_\text{rad}}
\def\Ein{E_\text{in}}
\def\Jin{J_\text{in}}

\newcommand{\TEOB}[1]{TEOBResumS{#1}}
\newcommand{\dali}{{Dal\'i}}

\def\eg{\textit{e.g.}}

\usepackage{pifont} 

\usepackage{color}
\definecolor{cyan}{rgb}{0,0.9,0.9}
\definecolor{orange}{rgb}{0.9,0.5,0}
\definecolor{magenta}{rgb}{1,0,1}
\definecolor{purple}{rgb}{0.8,0.4,0.8}
\definecolor{gray}{rgb}{0.8242,0.8242,0.8242}
\definecolor{light-gray}{gray}{0.95}

\begin{document}

\title{Gravitational scattering of two neutron stars}

\author{Joan \surname{Fontbuté}$^1$\orcid{0009-0004-7893-7386}}\email{joan.fontbute@uni-jena.de}
\author{Sebastiano \surname{Bernuzzi}$^1$\orcid{0000-0002-2334-0935}}\email{sebastiano.bernuzzi@uni-jena.de}
\author{Piero \surname{Rettegno}$^2$\orcid{0000-0001-8088-3517}}
\author{Simone \surname{Albanesi}$^{1,3}$\orcid{0000-0001-7345-4415}}
\author{Wolfgang \surname{Tichy}$^4$\orcid{0000-0002-8707-754X}}

\affiliation{${}^1$Theoretisch-Physikalisches Institut, Friedrich-Schiller-Universit{\"a}t Jena, 07743, Jena, Germany}
\affiliation{${}^2$Dipartimento di Fisica e Geologia, Università di Perugia,
	INFN Sezione di Perugia, Via Pascoli, I-06123 Perugia, Italy}
\affiliation{${}^3$INFN sezione di Torino, Torino, 10125, Italy}
\affiliation{${}^4$Department of Physics, Florida Atlantic University, Boca Raton, Florida 33431, USA}

\date{\today}

\begin{abstract}
  We present the first numerical relativity simulations of the gravitational scattering of two neutron stars.
  Constraint-satisfying initial data for two equal-mass nonspinning sequences are constructed at fixed energy and various initial angular momenta (impact parameter) and evolved with Einstein equations through the scattering process.
  The strong-field scattering dynamics are explored up to scattering angles of $220^\circ$ and the threshold of dynamical captures.   The transition to bound orbits is aided by significant mass ejecta up to baryon mass ${\sim}0.1\Mo$.
  A quantitative comparison with predictions of the scattering angle from state-of-the-art effective-one-body and post-Minkowskian calculations indicates quantitative agreement for large initial angular momenta although significant discrepancies in the tidal contribution emerge toward the capture threshold.
  Gravitational waveforms and radiated energy are in qualitative agreement with the analogous black hole problem and state-of-the-art effective-one-body predictions. Toward the capture threshold waveforms from scattering dynamics carry a strong imprint of matter effects, including the stars' $f$-mode excitations during the close encounter.
  Overall, our simulations open a new avenue to study tidal interactions in the relativistic two-body problem.
\end{abstract}


\maketitle

Numerical relativity (NR) plays a key role in the solution of the two-body problem in general relativity (GR). Simulations have been decisive to identify the inspiral-merger-ringdown process for compact binaries in bound orbits \cite{Pretorius:2006tp,Baker:2005vv,Campanelli:2005dd} and for the development of waveform templates for gravitational wave (GW) astronomy \cite{LIGOScientific:2016aoc}. %
NR calculations of the scattering process of two black holes were pioneered in Refs.~\cite{Shibata:2008rq,Sperhake:2009jz,Gold:2012tk,Sperhake:2012me,Damour:2014afa}
and recently extended to high energies, spinning and charged systems~\cite{Hopper:2022rwo,Rettegno:2023ghr,Long:2024ltn,Albanesi:2024xus,Swain:2024ngs,Smith:2024vka,Smith:2024nwq,Fontbute:2024amb,Albanesi:2025txj}.
Scattering simulations provide us with a unique access to the strong-field (high-velocity/frequency) regime of the motion. Therefore, advances in gravitational scattering simulations can further support the design of accurate GW models suitable for high-precision physics with future detectors~\cite{Punturo:2010zz,LISA:2017pwj,Reitze:2019iox,LISAConsortiumWaveformWorkingGroup:2023arg,Abac:2025saz}.

The effective-one-body (EOB) formalism is the contemporary analytical framework for the description of dynamics and waveforms from compact binaries~\cite{Buonanno:1998gg,Buonanno:2000ef}. State-of-the-art EOB models build on post-Newtonian (PN) results and, for bound orbits, achieve a complete description of the coalescence process using analytical resummations and NR information, \eg~\cite{Nagar:2021xnh,Nagar:2024oyk,Pompili:2023tna,Gamboa:2024hli,Albanesi:2025txj}. Some of these EOB models also deliver predictions for the motion of black holes in open (hyperbolic) orbits \cite{Damour:2014afa}, in spite of incorporating information from weak-field (slow motion) PN and bound (mostly circular) NR data. Recent work has shown that (PN-based) EOB models are able to faithfully predict the scattering angle for black holes with moderate initial energies up to $\Ein/M\lesssim 1.02$ even close to the bound-unbound transition~\cite{Nagar:2021xnh,Hopper:2022rwo,Nagar:2024oyk,Albanesi:2025txj}.

Gravitational scattering of compact bodies 
in GR recently gained significant interest because of advances in high-order post-Minkowskian (PM) calculations. Such weak-field approximations can be used to analytically describe unbound compact binary systems using a multitude of different approaches, generally inspired by quantum computations, such as
effective field theories \cite{Kalin:2020mvi,Kalin:2020fhe,Dlapa:2021vgp,Dlapa:2022lmu}, scattering amplitudes ~\cite{Vines:2017hyw,Guevara:2019fsj,Bern:2019nnu,Bern:2021dqo,Bern:2021yeh,Manohar:2022dea}, 
eikonalization~\cite{KoemansCollado:2019ggb,DiVecchia:2019kta,DiVecchia:2021bdo,DiVecchia:2021ndb}
and worldline field theory~\cite{Bini:2022wrq,Bini:2022enm,Jakobsen:2023ndj,Driesse:2024xad}. These computations have also been extended to take into account tidal deformations of extended bodies~\cite{Bini:2020flp,Bern:2020uwk,Cheung:2020sdj,Kalin:2020lmz,Jakobsen:2023pvx}.
Leading-order (LO) PM waveforms have been computed in Ref.~\cite{Kovacs:1977uw} and have recently been extended to next-to-leading order (NLO)~\cite{Brandhuber:2023hhy,Georgoudis:2023lgf}.
PM and NR scattering angles are compared in Refs.~\cite{Khalil:2022ylj,Damour:2022ybd,Rettegno:2023ghr,Buonanno:2024vkx}, while waveform comparisons face more issues because of both their numerical extraction (see \eg~Ref.~\cite{Albanesi:2024xus}) and the analytical complexity of PM results. 
EOB models that exploit PM information are currently under investigation \cite{Damour:2016gwp,Buonanno:2024byg,Damour:2025uka}.

In this Letter we present the first NR results for the scattering of two neutron stars.
The numerical solution of the scattering problem is obtained by solving the 3+1 Einstein equations as a Cauchy problem with suitable constraint-satisfying initial data.
Constraint equations are solved in the extended conformal thin sandwich formulation \cite{Baumgarte:1997eg,Pfeiffer:2002iy,Isenberg:2007zg} with a conformally flat 3-metric using the approach described in \cite{Moldenhauer:2014yaa} as implemented in the \texttt{SGRID} code \cite{Tichy:2011gw,Tichy:2012rp,Dietrich:2015pxa,Tichy:2019ouu}. While \texttt{SGRID} was developed for constructing binaries in bound orbits, hyperbolic orbits can be obtained by specifying the coordinate radial and angular velocity (algorithmic parameters) to their Newtonian values and fixing them during the elliptic solve by only adjusting the center of mass while keeping the total Arnowitt-Deser-Misner (ADM) momentum component $P^y_{\rm ADM} = 0$.
Sequences of two equal-mass nonspinning neutron stars (NS-NS) data with mass $M=2\times1.4\Mo$ (at infinite separation) are generated at fixed ADM energy $\Ein/M\simeq1.034$ and different angular momenta $\Jin$. The latter is related to the NR impact parameter $b$ and momentum by $\Jin = D|p_y| = |\vec{p}| b$, where $D=100M$ is the initial coordinate separation. The corresponding typical individual velocities are $v\sim2|\vec{p}|/M\lesssim0.27$. The fluid is initially assumed in stationary equilibrium and irrotational. We consider two sequences of scattering configurations described by piecewise polytropic equations of state (EOS) fitting the SLy (``soft'') and M1Sb (``stiff'') models \cite{Read:2008iy}. Tidal interactions are parametrized at leading order by the quadrupolar tidal polarizability constant~\cite{Damour:2009wj,Damour:2012yf} $\kappa_2^T=57$ ($236$) for the SLy (MS1b) configurations. 
Initial data parameters are summarized in Table.~\ref{tab:res}. 
Note that we have also extensively explored the evolution of superposed Tolman-Oppenheimer-Volkoff (nonconformally flat and constraint violating) data, but found that the evolution of those data is not sufficient to obtain the quantitative results presented below. In this Letter we use geometrized units $c=G=1$. 

Initial data are evolved with the free-evolution Z4c scheme \cite{Bernuzzi:2009ex,Hilditch:2012fp} and moving puncture gauge \cite{Alcubierre:2004hr,Baker:2006ha,Campanelli:2005dd}, coupled to relativistic hydrodynamics. For each NS-NS simulation, we also solve the corresponding black hole scattering problem (BH-BH) using Bowen-York puncture data~\cite{Bowen:1980yu,Brandt:1997tf,Ansorg:2004ds}.
We employ the \texttt{BAM} code \cite{Brugmann:2008zz,Thierfelder:2011yi} in which the computational domain is discretized on a hierarchy of Eulerian adaptive Cartesian grids. Specifically, we use seven refinement levels out of which four are moving and regridding with the star trackers (minimum of the lapse function). The finest boxes cover the stars entirely with a linear grid spacing $h_{7}=0.25,0.167$ and $0.125\Mo$ and $n=64,96$ and $128$ points in each direction. Simulations at multiple resolutions are performed to assess the impact of truncation errors. The outer boundary is located at coordinates $x^i_{\rm max}=-x^i_{\rm min}=1536\Mo$ and bitant symmetry (reflection through the Cartesian $z=x^3$ axis) is enforced. The moving puncture equations and parameters are identical to those in~\cite{Hilditch:2012fp}.
Metric derivatives are discretized with fourth-order finite difference operators; finite difference shock-capturing methods are used for hydrodynamics with weighted-essentially-nonoscillatory reconstruction \cite{Bernuzzi:2012ci}. Time evolution is performed with the method of lines and the fourth-order Runge-Kutta explicit scheme with a Courant-Friedrich-Lewy parameter of $0.25$.

We analyze the dynamics using gauge invariant quantities reported in Table.~\ref{tab:res}.
The scattering angle $\chi^{\rm NR}$ for each simulation sequence is computed from the star trackers following the polynomial extrapolation procedure used in previous work on black hole scattering, \eg~\cite{Damour:2014afa}; uncertainties are evaluated by summing in quadrature the independent uncertainties due to grid resolution and to the polynomial extrapolation.
GWs and the radiated energy/angular momentum are calculated from the GW modes (see below).
The reported ejecta mass is the baryon mass of the unbound fluid elements.
The binding energy of the system is computed as $E_b= \Ein - E_{\rm rad} - E_{\rm ejecta} - M$, where $E_{\rm ejecta}$ is a contribution originating from the mass ejecta. When the latter are relevant, $E_b$ is estimated from the ADM energy integrals at the GW extraction spheres assuming the balance law.
Error estimates are presented in parentheses of Table.~\ref{tab:res}; they are obtained by summing in quadrature truncation errors and finite extraction of GW modes and ADM integrals, see \eg~\cite{Bernuzzi:2012ci,Fontbute:2025ixd}.

\begin{table}[t]
  \centering    
  \caption{Initial data and simulation results after the first encounter.
    All configurations have $M=2\times1.4\Mo$ and $\Ein/M\simeq1.034$.
    Baryon mass for SLy (MS1b) is $M_b\sim1.56\Mo$ ($1.53\Mo$). Other quantities are defined in the text.}
  \scalebox{0.82}{
  \begin{tabular}{ccccccc}        
    \hline \hline
    EOS & $\Jin/M^2$ & $\Egw/M$ & $\Jgw/M^2$ & $E_b/M$ & $\chi^{\rm NR}$ [$^\circ$] & $M_b^{\rm ejecta}$ [$\Mo$]\\     \hline
    SLy                  & 1.192(7)             & 0.0115(3)            & 0.117(2)             & -0.01(1)             & ---      & 0.088(3)               \\
    SLy                  & 1.206(7)             & 0.0109(2)            & 0.113(2)             & -0.005(8)            & ---      & 0.089(4)               \\
    SLy                  & 1.220(7)             & 0.0103(2)            & 0.109(1)             & 0.02(2)              & ---      & 0.088(5)               \\
    SLy                  & 1.248(8)             & 0.0087(2)            & 0.101(7)             & 0.020(6)             & 221(3)   & 0.096(4)               \\
    SLy                  & 1.290(8)             & 0.0069(2)            & 0.098(10)            & 0.023(4)             & 178.0(3) & 0.073(1)              \\
    SLy                  & 1.317(8)             & 0.0060(3)            & 0.10(2)              & 0.024(4)             & 159.2(2) & 0.053(2)             \\
    SLy                  & 1.387(9)             & 0.0040(2)            & 0.09(3)              & 0.027(3)             & 131.1(2) & 0.020(4)              \\
    SLy                  & 1.456(9)             & 0.00276(4)           & 0.050(8)             & 0.029(2)             & 113.6(4) & 0.004(4)             \\
    SLy                  & 1.53(1)              & 0.00204(4)           & 0.045(8)             & 0.03189(4)           & 101.4(4) & 0.001(4)           \\
    SLy                  & 1.59(1)              & 0.00154(4)           & 0.037(7)             & 0.03239(4)           & 92.0(4)  & $<10^{-3}$           \\
    SLy                  & 1.66(1)              & 0.00121(5)           & 0.03(1)              & 0.03272(5)           & 84.5(4)  & $<10^{-3}$           \\
    SLy                  & 1.80(1)              & 0.00082(9)           & 0.04(2)              & 0.03311(7)           & 72.9(5)  & $<10^{-3}$           \\
    SLy                  & 2.08(1)              & 0.0007(3)            & 0.09(8)              & 0.0332(3)            & 57.8(6)  & $<10^{-3}$           \\
    SLy                  & 2.50(1)              & 0.00025(9)           & 0.02(2)              & 0.03368(9)           & 45(1)    & $<10^{-3}$           \\
    SLy                  & 3.05(2)              & 0.00013(5)           & 0.009(7)             & 0.03380(5)           & 34(1)    & $<10^{-3}$           \\ \hline
    MS1b                 & 1.313(5)             & 0.0047(1)            & 0.09(2)              & 0.027(2)             & 162(1)*  & 0.076(7)              \\
    MS1b                 & 1.326(5)             & 0.0045(1)            & 0.09(2)              & 0.027(2)             & 159(1)*  & 0.073(9)              \\
    MS1b                 & 1.340(5)             & 0.0043(2)            & 0.09(2)              & 0.027(2)             & 155(1)*  & 0.071(7)              \\
    MS1b                 & 1.354(5)             & 0.0042(2)            & 0.09(3)              & 0.027(2)             & 151(1)*  & 0.066(5)             \\
    MS1b                 & 1.368(5)             & 0.0040(2)            & 0.09(4)              & 0.028(2)             & 147.2(6) & 0.063(7)              \\
    MS1b                 & 1.382(6)             & 0.0035(3)            & 0.07(9)              & 0.028(1)             & 141.4(6) & 0.06(1)              \\
    MS1b                 & 1.451(6)             & 0.00264(6)           & 0.06(1)              & 0.030(1)             & 120.4(8) & 0.03(7)             \\
    MS1b                 & 1.520(6)             & 0.00196(4)           & 0.043(8)             & 0.0308(9)            & 105.3(8) & 0.003(9)             \\
    MS1b                 & 1.589(6)             & 0.00150(3)           & 0.038(7)             & 0.0310(8)            & 94.8(5)  & $<10^{-3}$             \\
    MS1b                 & 1.658(7)             & 0.00118(6)           & 0.03(1)              & 0.031(3)             & 87(1)    & $<10^{-3}$             \\
    MS1b                 & 1.796(7)             & 0.00081(9)           & 0.03(2)              & 0.03312(9)           & 74.1(5)  & $<10^{-3}$           \\
    MS1b                 & 2.072(8)             & 0.0008(4)            & 0.1(1)               & 0.0331(4)            & 58.4(6)  & $<10^{-3}$           \\
    MS1b                 & 2.49(1)              & 0.00026(9)           & 0.02(2)              & 0.03367(9)           & 44.7(7)  & $<10^{-3}$           \\
    MS1b                 & 3.040(9)             & 0.00013(5)           & 0.009(7)             & 0.03380(5)           & 34.5(9)  & $<10^{-3}$           \\
    \hline \hline
  \end{tabular}
  }
 \label{tab:res}
\end{table}

\begin{figure*}[t!]
  \centering 
  \includegraphics[width=0.98\textwidth]{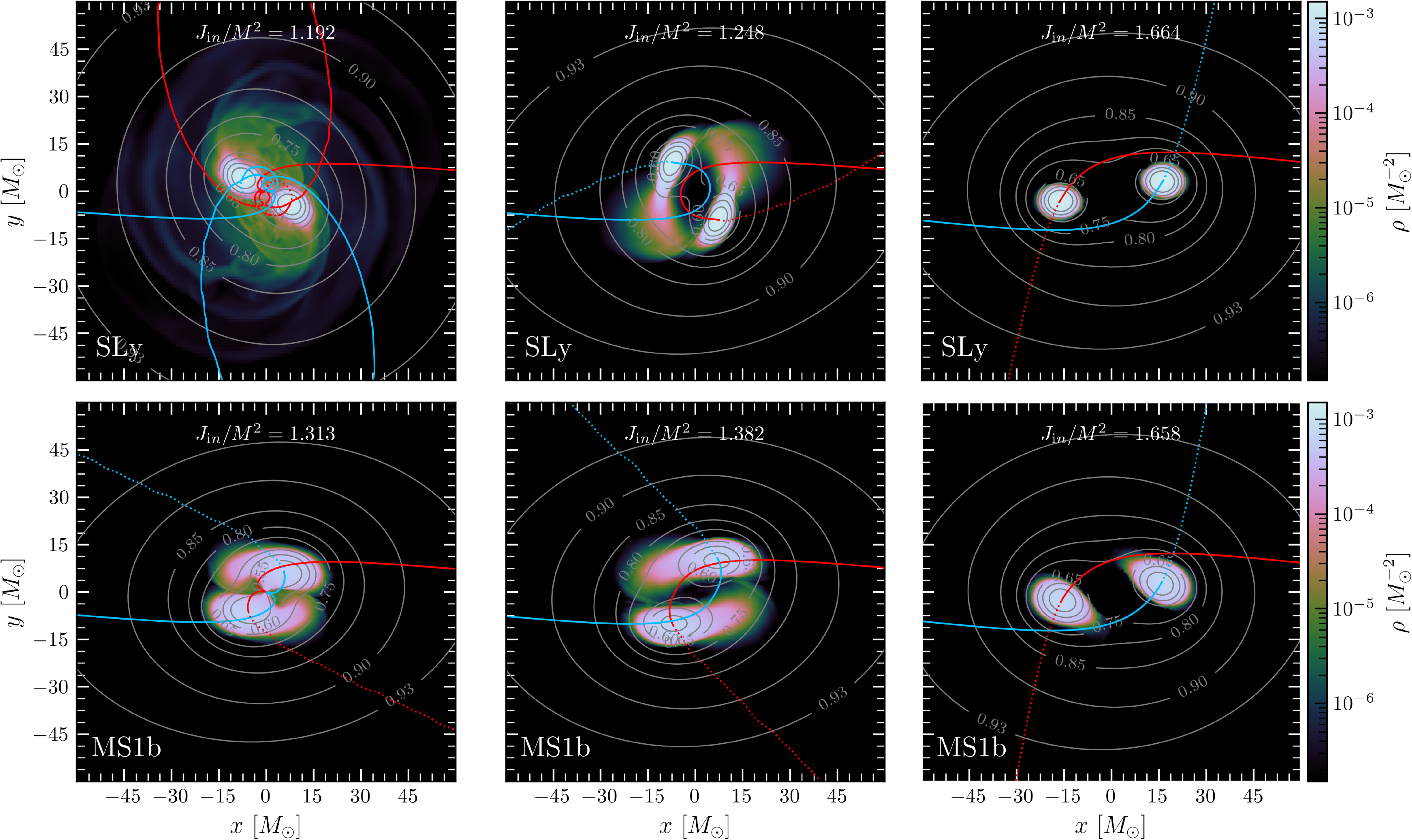}
  \caption{Snapshots of SLy simulations with $\Jin/M^2=1.192,1.248,1.664$ (upper row) and MS1b simulations with $\Jin/M^2=1.313,1.382,1.658$ (lower row) both at $\Ein/M\simeq1.034$. Each panel shows the rest-mass density (color map) and the 3-metric conformal factor (contours at $0.5, 0.55, 0.6, 0.65, 0.75, 0.8, 0.85, 0.9, 0.93$) on the orbital plane and shortly after the close encounter. The star trackers are plotted with red and blue solid lines up to the snapshots and discontinuous from then on. Smaller initial angular momenta result in scattering with a closer passage and eventually to a dynamical capture (merger).}
 \label{fig:dyn}
\end{figure*}
The phenomenology of the scattering process is illustrated in Fig.~\ref{fig:dyn}.
Configurations with progressively lower $\Jin$ (and fixed-energy $\Ein$) undergo quasielastic scattering with increasingly larger scattering angles.~\footnote{We also observe an increasingly larger star spin induced during the encounter; since quasilocal diagnostics were not enabled for these simulations we cannot quantify this effect.}
At sufficiently low $\Jin$ and stiff EOS, the stars may come in contact but still scatter apart from each other [cases marked with an asterisk (*) in Table.~\ref{tab:res}]. Below a critical threshold value $\Jin^{\rm thre}$ the dynamics become bound ($E_b<0$) due to both GWs emission and mass ejecta triggered by tidal torque. In these cases, the stars are captured after the first close encounter and merge. Dynamical mass ejection becomes significant already at $\Jin/M^2\lesssim1.52$ and up to ${\sim}0.1\Mo$ of unbound baryon mass are launched during the close encounter for $\Jin\sim\Jin^{\rm thre}$. These ejecta are about 1 order of magnitude larger or comparable to those from both circularized (highly eccentric but initially bound) and dynamical capture (initially unbound) mergers \cite{Radice:2016dwd, Papenfort:2018bjk, Chaurasia:2018zhg, Neuweiler:2025lte}, depending on the initial angular momentum. They crucially contribute to lower the binding energy to negative values, as illustrated by Table.~\ref{tab:res} (note that some of the material might fall back at larger timescale).
From the simulated sequence we estimate the threshold $\Jin^{\rm thre}/M^2\approx (1.25+1.22)/2=1.235$ ($\Jin^{\rm thre}/M^2\lesssim 1.31$) for the SLy (MS1b) case.

\begin{figure}[t!]
  \centering 
  \includegraphics[width=0.49\textwidth]{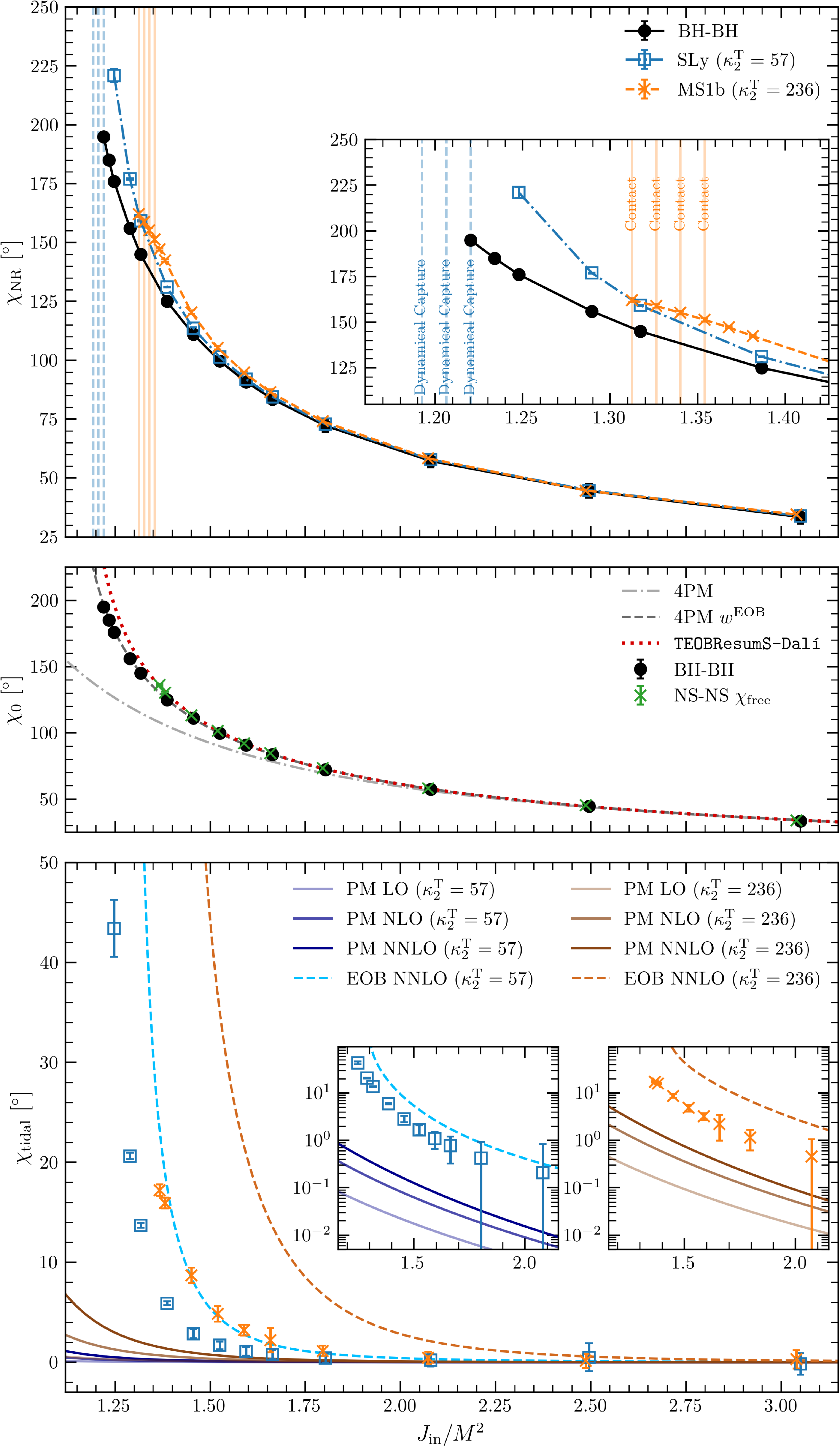}
  \caption{Scattering angle as a function of the initial angular momentum for the two simulation series. Data refer to the highest resolution set of simulations.     Top: total scattering angle $\chi^{\rm NR}$ for neutron stars and black holes with same initial energy $\Ein/M \simeq 1.034$ and angular momenta $\Jin/M^2$.
    Middle: scattering angle for the BH-BH sequence and comparison to the tidal-free term $\chi^{\rm NR}_{\rm free}$ from neutron star scattering (see main text) and to \texttt{\TEOB{-\dali}{}} and 4PM result in expanded or EOB-resummed form.
    Bottom: tidal contribution to the scattering angle (computed by subtracting the BH-BH data) and comparison with \texttt{\TEOB{-\dali}{}} (PN NNLO tides) and PM calculations up to NNLO.}
 \label{fig:scata}
\end{figure}

The scattering angle $\chi^{\rm NR}$ for each simulation sequence is shown in Fig.~\ref{fig:scata} (top panel). Our simulations probe scattering angles up to $\chi^{\rm NR}\sim 220^\circ$, after which the dynamical capture regime settles in. Significant deviations from the black hole case $\chi^{\rm NR}_0$ are visible for $\Jin/M^2\lesssim1.75$.

Numerical results are compared to analytical predictions for nonspinning configurations from the (PN-based) EOB model \texttt{\TEOB{-\dali}{}} \cite{Albanesi:2025txj} and from post-Minkowskian calculations up to 4PM, both in expanded and in EOB-resummed form \cite{Damour:2022ybd}.
Tidal interactions in \texttt{\TEOB{-\dali}{}} are modeled either using all available next-to-next-to-leading order (NNLO) PN information~\cite{Bernuzzi:2012ci,Gamba:2023mww} or the gravitational self-force resummation of that NNLO PN model~\cite{Bini:2012gu,Bernuzzi:2014owa,Akcay:2018yyh}. 
Tidal effects in PM are considered up to NNLO (4PM)~\cite{Bini:2020flp,Kalin:2020lmz,Jakobsen:2023hig} for the quadrupolar tides and NLO for the octupolar ones~\cite{Kalin:2020lmz} .~\footnote{At NNLO~\cite{Jakobsen:2023pvx} we neglected post-adiabatic tides and chose the stars' radii as length scale $R$.}
For a quantitative comparison, we compute the tidal contribution by subtracting the NR BH-BH data $\chi^{\rm NR}_{\rm tidal}=\chi^{\rm NR}-\chi^{\rm NR}_0$ .
The middle panel of Fig.~\ref{fig:scata} shows the BH-BH contribution $\chi^{\rm NR}_0$ and analytical predictions. \texttt{\TEOB{-\dali}{}} reproduces quite well NR data at the considered energy, as discussed also in \eg~\cite{Nagar:2021xnh,Albanesi:2025txj}. 
Post-Minkowskian calculations at 4PM reproduce $\chi^{\rm NR}_0$ only for $\Jin/M^2\gtrsim1.75$, while the accurate description of NR data requires a EOB resummation~\cite{Damour:2022ybd} (see also \cite{Buonanno:2024vkx}).
Assuming the approximate parametrization $\chi\approx\chi_{\rm free}+\kappa_2^T u$ with $u$ a function weakly dependent on tidal parameters, we calculate from the two simulated sequences the ``tidal-free'' term $\chi_{\rm free}^{\rm NR}=[\kappa_2^T({\rm SLy})\chi^{\rm NR}({\rm MS1b})-\kappa_2^T({\rm MS1b})\chi^{\rm NR}({\rm SLy})]/[\kappa_2^T({\rm SLy})-\kappa_2^T({\rm MS1b})]$~\cite{Damour:2009wj}. The middle panel of Fig.~\ref{fig:scata} shows that $\chi_{\rm free}^{\rm NR}$ is compatible with the BH-BH $\chi^{\rm NR}_0$. This demonstrates that $\chi^{\rm NR}$ from neutron star simulations consistently incorporates the point-mass contribution as in BH-BH simulations, and that $u$ is indeed an approximately EOS-insensitive (quasiuniversal) function. 

The bottom panel of Fig.~\ref{fig:scata} shows the tidal contribution $\chi^{\rm NR}_{\rm tidal}$ and the analytical predictions. The EOB NNLO tidal curve is in good agreement to the actual data only down to $\Jin/M^2\gtrsim1.6$ ($\Jin/M^2\gtrsim2$) for SLy (M1Sb), while it overestimates the data for lower angular momenta. We find even larger deviations with the gravitational self-force resummed tidal EOB potential (not shown in the figure). These EOB prescriptions for tidal interactions result in too strong tides for the accurate description of the scattering process.
Considering PM-expanded $\chi_{\rm tidal}$ we only find acceptable agreement at high initial angular momenta $\Jin/M^2\gtrsim1.75$ ($\Jin/M^2\gtrsim2$) for SLy (M1Sb). At lower $\Jin/M^2$, NNLO predictions considerably underestimate the tidal corrections to the scattering angle even when the mass ejecta effect is still negligible, specially the lower the tidal polarizability as for the SLy case.

 \begin{figure}[t]
   \centering 
   \includegraphics[width=0.49\textwidth]{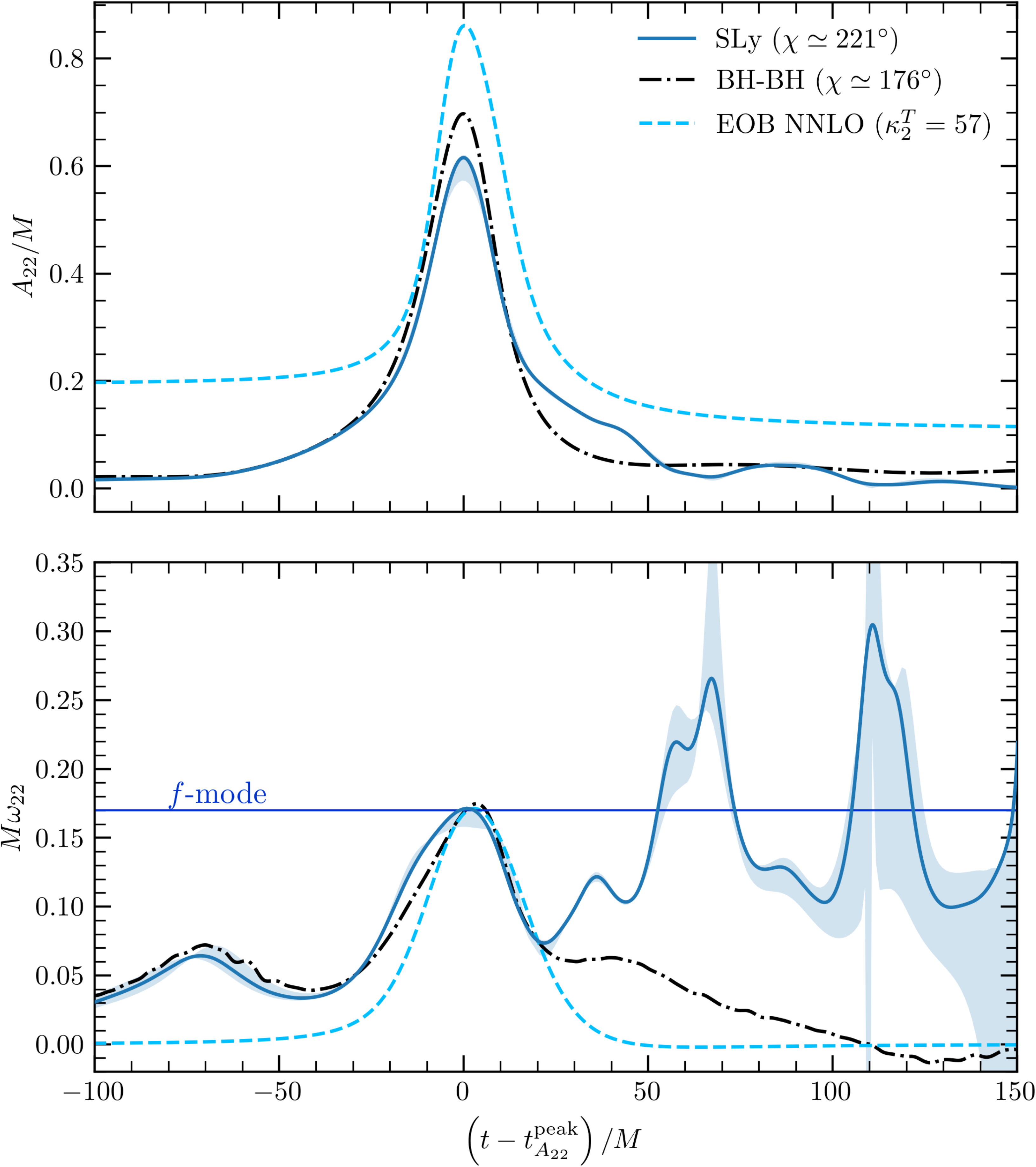}
   \caption{Gravitational waveform amplitude (top) and frequency (bottom) for the SLy NS-NS with $\Jin/M^2=1.248$ (solid blue, extracted at $R=400\Mo$), for the BH-BH case with same $\Jin$ (black, extracted at $R=100\Mo$) and for EOB NNLO tides (dashed light blue).
   The EOB initial angular momentum is chosen such that the peak of the EOB frequency matches the NS-NS one. The blue shaded area corresponds to the numerical uncertainty for the NS-NS case linked to resolution and extraction radius.}
  \label{fig:gw}
 \end{figure}

Gravitational waveforms are computed from Weyl's $\psi_4$ pseudoscalar, \eg~\cite{Campanelli:2005ia,Brugmann:2008zz} on extraction spheres of coordinate isotropic radii $R=(x_ix^i)^{1/2}=300,400,500,600,700\Mo$.
Waveforms are found convergent with resolution at approximately third order around the GW burst and consistently in the different runs; convergence drops to first order after the encounter.
Multipoles of the GW strain are calculated by decomposing $\psi_4$ in spin-weighted spherical harmonics with spin $s=-2$ and integrating the relation $\ddot{h}=\psi_4\big|_{R=\infty}$, 
where we use the perturbative approach described in \cite{Nakano:2015pta} to extrapolate the $\psi_4$ modes to spatial infinity. 
Note that the strain is obtained from the extrapolated Weyl pseudoscalar via a 
double fixed-frequency integration~\cite{Reisswig:2010di}, which cuts off the low-frequency 
part of the signal. As a consequence, the amplitude of the waveform approaches zero before and after the close encounter, whereas the waveform is expected to have a nonzero constant amplitude 
due to linear memory effects~\cite{Zeldovich:1974gvh,Braginsky:1985vlg}. 
These effects are instead included in the EOB waveform at Newtonian order~\cite{Chiaramello:2020ehz}.

Waveforms from configurations with $\Jin/M^2\gtrsim 1.55$ broadly agree with the BH-BH case.
The most interesting features in the scattering waveforms are found in configurations approaching $\Jin^{\rm thre}$ from above; Fig.~\ref{fig:gw} shows a representative example. 
The peak amplitude (top panel) of the $\ell=m=2$ mode is significantly reduced with respect to the BH-BH case although, for a given $\Jin$, tidal effects accelerating the transition to bound orbits are expected to generate a larger amplitude than for black holes. We attribute this behavior to massive ejecta observed in these scatterings. Comparison to the amplitude of \texttt{\TEOB{-\dali}{}} (NNLO) waveforms, which do not contain ejecta effects, supports the interpretation. This effect is mirrored in the radiated energy which can be reduced up to a factor ${\sim}4$, see Table.~\ref{tab:res}. Note again that the radiated GW energy during the encounter alone is insufficient to drive the dynamics to bound orbits.
The peak frequencies (bottom panel) are found up to $M\omega_{22}\sim 0.2$ (${\sim}2$ kHz) and agree rather well with the BH-BH and EOB NNLO predictions. 
SLy simulations approaching the dynamical capture threshold are also characterized by a postpeak signature of the star $f$-mode oscillations excited by the close encounter. This is similar to what observed in dynamical captures \cite{Gold:2011df}, although for scattering the $f$-mode contribution is less persistent and has weaker amplitude.

In summary, the computations presented here are a first step toward bridging NR computations to modern analytical approaches to the relativistic two-body problem. Numerical data can guide and motivate higher-order perturbative calculations in post-Minkowskian gravity and the study of resummation techniques, \eg~\cite{Damour:2022ybd}. We anticipate that scattering simulations will inform the next generation of EOB models and improve the description of tidal interactions in the high-frequency regime.  Such improved understanding is crucial to obtain robust constraints on extreme matter from observations~\cite{Damour:2012yf,TheLIGOScientific:2017qsa,De:2018uhw,LIGOScientific:2018hze}. Furthermore, our computations predict very massive neutron rich ejecta that would shine as kilonova counterparts to the GW~\cite{Rosswog:2012wb}. These counterparts, together with observed $r$-process element abundancies, could provide a unique way of constraining astrophysical rates of neutron star scattering events. Work in these directions is ongoing.
\\

  The authors thank Thibault Damour, Rossella Gamba, David Radice, Radu Roiban, and Malte Schulze for input and discussions.
  SB thanks the Institute for Gravitation and the Cosmos at Penn State for
  the hospitality and support during his visit in December 2023. %
  SB thanks the Institut des Hautes Études Scientifiques for hospitality and support during his visit in February 2024. 
  JF and SB acknowledge support by the EU Horizon under ERC Consolidator Grant, no. InspiReM-101043372.
  PR acknowledges support from the Italian Ministry of University and Research (MUR) via the PRIN 2022ZHYFA2, GRavitational wavEform models for coalescing compAct binaries with eccenTricity (GREAT) and through the program
  “Dipartimenti di Eccellenza 2018-2022” (Grant SUPER-C).
  PR also acknowledges support from “Fondo di Ricerca d’Ateneo” of the University of Perugia.
  SA and SB acknowledge support from the Deutsche Forschungsgemeinschaft (DFG) project ``GROOVHY'' 
  (BE 6301/5-1 Projektnummer: 523180871).
  WT was supported by the National Science Foundation under grants PHY-2136036
  and PHY-2408903. 

Simulations were performed on SuperMUC-NG at the Leibniz-Rechenzentrum
(LRZ) Munich and and on the national HPE Apollo Hawk at the High
Performance Computing Center Stuttgart (HLRS). 
The authors acknowledge the Gauss Centre for Supercomputing
e.V. (\url{www.gauss-centre.eu}) for funding this project by providing
computing time on the GCS Supercomputer SuperMUC-NG at LRZ
(allocations {\tt pn36go}, {\tt pn36jo} and {\tt pn68wi}). The authors
acknowledge HLRS for funding this project by providing access to the
supercomputer HPE Apollo Hawk under the grant number INTRHYGUE/44215
and MAGNETIST/44288. 
Computations were also performed on the ARA cluster at Friedrich
Schiller University Jena and on the {\tt Tullio} INFN cluster at INFN
Turin. The ARA cluster is funded in part by DFG grants INST 275/334-1
FUGG and INST 275/363-1 FUGG, and ERC Starting Grant, grant agreement
no. BinGraSp-714626.


\begin{thebibliography}{103}%
\makeatletter
\providecommand \@ifxundefined [1]{%
 \@ifx{#1\undefined}
}%
\providecommand \@ifnum [1]{%
 \ifnum #1\expandafter \@firstoftwo
 \else \expandafter \@secondoftwo
 \fi
}%
\providecommand \@ifx [1]{%
 \ifx #1\expandafter \@firstoftwo
 \else \expandafter \@secondoftwo
 \fi
}%
\providecommand \natexlab [1]{#1}%
\providecommand \enquote  [1]{``#1''}%
\providecommand \bibnamefont  [1]{#1}%
\providecommand \bibfnamefont [1]{#1}%
\providecommand \citenamefont [1]{#1}%
\providecommand \href@noop [0]{\@secondoftwo}%
\providecommand \href [0]{\begingroup \@sanitize@url \@href}%
\providecommand \@href[1]{\@@startlink{#1}\@@href}%
\providecommand \@@href[1]{\endgroup#1\@@endlink}%
\providecommand \@sanitize@url [0]{\catcode `\\12\catcode `\$12\catcode
  `\&12\catcode `\#12\catcode `\^12\catcode `\_12\catcode `\%12\relax}%
\providecommand \@@startlink[1]{}%
\providecommand \@@endlink[0]{}%
\providecommand \url  [0]{\begingroup\@sanitize@url \@url }%
\providecommand \@url [1]{\endgroup\@href {#1}{\urlprefix }}%
\providecommand \urlprefix  [0]{URL }%
\providecommand \Eprint [0]{\href }%
\providecommand \doibase [0]{http://dx.doi.org/}%
\providecommand \selectlanguage [0]{\@gobble}%
\providecommand \bibinfo  [0]{\@secondoftwo}%
\providecommand \bibfield  [0]{\@secondoftwo}%
\providecommand \translation [1]{[#1]}%
\providecommand \BibitemOpen [0]{}%
\providecommand \bibitemStop [0]{}%
\providecommand \bibitemNoStop [0]{.\EOS\space}%
\providecommand \EOS [0]{\spacefactor3000\relax}%
\providecommand \BibitemShut  [1]{\csname bibitem#1\endcsname}%
\let\auto@bib@innerbib\@empty
\bibitem [{\citenamefont {Pretorius}(2006)}]{Pretorius:2006tp}%
  \BibitemOpen
  \bibfield  {author} {\bibinfo {author} {\bibfnamefont {F.}~\bibnamefont
  {Pretorius}},\ }\href {\doibase 10.1088/0264-9381/23/16/S13} {\bibfield
  {journal} {\bibinfo  {journal} {Class.Quant.Grav.}\ }\textbf {\bibinfo
  {volume} {23}},\ \bibinfo {pages} {S529} (\bibinfo {year} {2006})},\ \Eprint
  {http://arxiv.org/abs/gr-qc/0602115} {arXiv:gr-qc/0602115 [gr-qc]}
  \BibitemShut {NoStop}%
\bibitem [{\citenamefont {Baker}\ \emph {et~al.}(2006)\citenamefont {Baker},
  \citenamefont {Centrella}, \citenamefont {Choi}, \citenamefont {Koppitz},\
  and\ \citenamefont {van Meter}}]{Baker:2005vv}%
  \BibitemOpen
  \bibfield  {author} {\bibinfo {author} {\bibfnamefont {J.~G.}\ \bibnamefont
  {Baker}}, \bibinfo {author} {\bibfnamefont {J.}~\bibnamefont {Centrella}},
  \bibinfo {author} {\bibfnamefont {D.-I.}\ \bibnamefont {Choi}}, \bibinfo
  {author} {\bibfnamefont {M.}~\bibnamefont {Koppitz}}, \ and\ \bibinfo
  {author} {\bibfnamefont {J.}~\bibnamefont {van Meter}},\ }\href {\doibase
  10.1103/PhysRevLett.96.111102} {\bibfield  {journal} {\bibinfo  {journal}
  {Phys. Rev. Lett.}\ }\textbf {\bibinfo {volume} {96}},\ \bibinfo {pages}
  {111102} (\bibinfo {year} {2006})},\ \Eprint
  {http://arxiv.org/abs/gr-qc/0511103} {arXiv:gr-qc/0511103} \BibitemShut
  {NoStop}%
\bibitem [{\citenamefont {Campanelli}\ \emph
  {et~al.}(2006{\natexlab{a}})\citenamefont {Campanelli}, \citenamefont
  {Lousto}, \citenamefont {Marronetti},\ and\ \citenamefont
  {Zlochower}}]{Campanelli:2005dd}%
  \BibitemOpen
  \bibfield  {author} {\bibinfo {author} {\bibfnamefont {M.}~\bibnamefont
  {Campanelli}}, \bibinfo {author} {\bibfnamefont {C.~O.}\ \bibnamefont
  {Lousto}}, \bibinfo {author} {\bibfnamefont {P.}~\bibnamefont {Marronetti}},
  \ and\ \bibinfo {author} {\bibfnamefont {Y.}~\bibnamefont {Zlochower}},\
  }\href {\doibase 10.1103/PhysRevLett.96.111101} {\bibfield  {journal}
  {\bibinfo  {journal} {Phys. Rev. Lett.}\ }\textbf {\bibinfo {volume} {96}},\
  \bibinfo {pages} {111101} (\bibinfo {year} {2006}{\natexlab{a}})},\ \Eprint
  {http://arxiv.org/abs/gr-qc/0511048} {arXiv:gr-qc/0511048} \BibitemShut
  {NoStop}%
\bibitem [{\citenamefont {Abbott}\ \emph {et~al.}(2016)\citenamefont {Abbott}
  \emph {et~al.}}]{LIGOScientific:2016aoc}%
  \BibitemOpen
  \bibfield  {author} {\bibinfo {author} {\bibfnamefont {B.~P.}\ \bibnamefont
  {Abbott}} \emph {et~al.} (\bibinfo {collaboration} {LIGO Scientific,
  Virgo}),\ }\href {\doibase 10.1103/PhysRevLett.116.061102} {\bibfield
  {journal} {\bibinfo  {journal} {Phys. Rev. Lett.}\ }\textbf {\bibinfo
  {volume} {116}},\ \bibinfo {pages} {061102} (\bibinfo {year} {2016})},\
  \Eprint {http://arxiv.org/abs/1602.03837} {arXiv:1602.03837 [gr-qc]}
  \BibitemShut {NoStop}%
\bibitem [{\citenamefont {Shibata}\ \emph {et~al.}(2008)\citenamefont
  {Shibata}, \citenamefont {Okawa},\ and\ \citenamefont
  {Yamamoto}}]{Shibata:2008rq}%
  \BibitemOpen
  \bibfield  {author} {\bibinfo {author} {\bibfnamefont {M.}~\bibnamefont
  {Shibata}}, \bibinfo {author} {\bibfnamefont {H.}~\bibnamefont {Okawa}}, \
  and\ \bibinfo {author} {\bibfnamefont {T.}~\bibnamefont {Yamamoto}},\ }\href
  {\doibase 10.1103/PhysRevD.78.101501} {\bibfield  {journal} {\bibinfo
  {journal} {Phys. Rev. D}\ }\textbf {\bibinfo {volume} {78}},\ \bibinfo
  {pages} {101501} (\bibinfo {year} {2008})},\ \Eprint
  {http://arxiv.org/abs/0810.4735} {arXiv:0810.4735 [gr-qc]} \BibitemShut
  {NoStop}%
\bibitem [{\citenamefont {Sperhake}\ \emph {et~al.}(2009)\citenamefont
  {Sperhake}, \citenamefont {Cardoso}, \citenamefont {Pretorius}, \citenamefont
  {Berti}, \citenamefont {Hinderer} \emph {et~al.}}]{Sperhake:2009jz}%
  \BibitemOpen
  \bibfield  {author} {\bibinfo {author} {\bibfnamefont {U.}~\bibnamefont
  {Sperhake}}, \bibinfo {author} {\bibfnamefont {V.}~\bibnamefont {Cardoso}},
  \bibinfo {author} {\bibfnamefont {F.}~\bibnamefont {Pretorius}}, \bibinfo
  {author} {\bibfnamefont {E.}~\bibnamefont {Berti}}, \bibinfo {author}
  {\bibfnamefont {T.}~\bibnamefont {Hinderer}},  \emph {et~al.},\ }\href
  {\doibase 10.1103/PhysRevLett.103.131102} {\bibfield  {journal} {\bibinfo
  {journal} {Phys.Rev.Lett.}\ }\textbf {\bibinfo {volume} {103}},\ \bibinfo
  {pages} {131102} (\bibinfo {year} {2009})},\ \Eprint
  {http://arxiv.org/abs/0907.1252} {arXiv:0907.1252 [gr-qc]} \BibitemShut
  {NoStop}%
\bibitem [{\citenamefont {Gold}\ and\ \citenamefont
  {Br{\"u}gmann}(2013)}]{Gold:2012tk}%
  \BibitemOpen
  \bibfield  {author} {\bibinfo {author} {\bibfnamefont {R.}~\bibnamefont
  {Gold}}\ and\ \bibinfo {author} {\bibfnamefont {B.}~\bibnamefont
  {Br{\"u}gmann}},\ }\href {\doibase 10.1103/PhysRevD.88.064051} {\bibfield
  {journal} {\bibinfo  {journal} {Phys. Rev.}\ }\textbf {\bibinfo {volume}
  {D88}},\ \bibinfo {pages} {064051} (\bibinfo {year} {2013})},\ \Eprint
  {http://arxiv.org/abs/1209.4085} {arXiv:1209.4085 [gr-qc]} \BibitemShut
  {NoStop}%
\bibitem [{\citenamefont {Sperhake}\ \emph {et~al.}(2013)\citenamefont
  {Sperhake}, \citenamefont {Berti}, \citenamefont {Cardoso},\ and\
  \citenamefont {Pretorius}}]{Sperhake:2012me}%
  \BibitemOpen
  \bibfield  {author} {\bibinfo {author} {\bibfnamefont {U.}~\bibnamefont
  {Sperhake}}, \bibinfo {author} {\bibfnamefont {E.}~\bibnamefont {Berti}},
  \bibinfo {author} {\bibfnamefont {V.}~\bibnamefont {Cardoso}}, \ and\
  \bibinfo {author} {\bibfnamefont {F.}~\bibnamefont {Pretorius}},\ }\href
  {\doibase 10.1103/PhysRevLett.111.041101} {\bibfield  {journal} {\bibinfo
  {journal} {Phys. Rev. Lett.}\ }\textbf {\bibinfo {volume} {111}},\ \bibinfo
  {pages} {041101} (\bibinfo {year} {2013})},\ \Eprint
  {http://arxiv.org/abs/1211.6114} {arXiv:1211.6114 [gr-qc]} \BibitemShut
  {NoStop}%
\bibitem [{\citenamefont {Damour}\ \emph {et~al.}(2014)\citenamefont {Damour},
  \citenamefont {Guercilena}, \citenamefont {Hinder}, \citenamefont {Hopper},
  \citenamefont {Nagar},\ and\ \citenamefont {Rezzolla}}]{Damour:2014afa}%
  \BibitemOpen
  \bibfield  {author} {\bibinfo {author} {\bibfnamefont {T.}~\bibnamefont
  {Damour}}, \bibinfo {author} {\bibfnamefont {F.}~\bibnamefont {Guercilena}},
  \bibinfo {author} {\bibfnamefont {I.}~\bibnamefont {Hinder}}, \bibinfo
  {author} {\bibfnamefont {S.}~\bibnamefont {Hopper}}, \bibinfo {author}
  {\bibfnamefont {A.}~\bibnamefont {Nagar}}, \ and\ \bibinfo {author}
  {\bibfnamefont {L.}~\bibnamefont {Rezzolla}},\ }\href {\doibase
  10.1103/PhysRevD.89.081503} {\bibfield  {journal} {\bibinfo  {journal} {Phys.
  Rev. D}\ }\textbf {\bibinfo {volume} {89}},\ \bibinfo {pages} {081503}
  (\bibinfo {year} {2014})},\ \Eprint {http://arxiv.org/abs/1402.7307}
  {arXiv:1402.7307 [gr-qc]} \BibitemShut {NoStop}%
\bibitem [{\citenamefont {Hopper}\ \emph {et~al.}(2023)\citenamefont {Hopper},
  \citenamefont {Nagar},\ and\ \citenamefont {Rettegno}}]{Hopper:2022rwo}%
  \BibitemOpen
  \bibfield  {author} {\bibinfo {author} {\bibfnamefont {S.}~\bibnamefont
  {Hopper}}, \bibinfo {author} {\bibfnamefont {A.}~\bibnamefont {Nagar}}, \
  and\ \bibinfo {author} {\bibfnamefont {P.}~\bibnamefont {Rettegno}},\ }\href
  {\doibase 10.1103/PhysRevD.107.124034} {\bibfield  {journal} {\bibinfo
  {journal} {Phys. Rev. D}\ }\textbf {\bibinfo {volume} {107}},\ \bibinfo
  {pages} {124034} (\bibinfo {year} {2023})},\ \Eprint
  {http://arxiv.org/abs/2204.10299} {arXiv:2204.10299 [gr-qc]} \BibitemShut
  {NoStop}%
\bibitem [{\citenamefont {Rettegno}\ \emph {et~al.}(2023)\citenamefont
  {Rettegno}, \citenamefont {Pratten}, \citenamefont {Thomas}, \citenamefont
  {Schmidt},\ and\ \citenamefont {Damour}}]{Rettegno:2023ghr}%
  \BibitemOpen
  \bibfield  {author} {\bibinfo {author} {\bibfnamefont {P.}~\bibnamefont
  {Rettegno}}, \bibinfo {author} {\bibfnamefont {G.}~\bibnamefont {Pratten}},
  \bibinfo {author} {\bibfnamefont {L.~M.}\ \bibnamefont {Thomas}}, \bibinfo
  {author} {\bibfnamefont {P.}~\bibnamefont {Schmidt}}, \ and\ \bibinfo
  {author} {\bibfnamefont {T.}~\bibnamefont {Damour}},\ }\href {\doibase
  10.1103/PhysRevD.108.124016} {\bibfield  {journal} {\bibinfo  {journal}
  {Phys. Rev. D}\ }\textbf {\bibinfo {volume} {108}},\ \bibinfo {pages}
  {124016} (\bibinfo {year} {2023})},\ \Eprint
  {http://arxiv.org/abs/2307.06999} {arXiv:2307.06999 [gr-qc]} \BibitemShut
  {NoStop}%
\bibitem [{\citenamefont {Long}\ \emph {et~al.}(2024)\citenamefont {Long},
  \citenamefont {Whittall},\ and\ \citenamefont {Barack}}]{Long:2024ltn}%
  \BibitemOpen
  \bibfield  {author} {\bibinfo {author} {\bibfnamefont {O.}~\bibnamefont
  {Long}}, \bibinfo {author} {\bibfnamefont {C.}~\bibnamefont {Whittall}}, \
  and\ \bibinfo {author} {\bibfnamefont {L.}~\bibnamefont {Barack}},\ }\href
  {\doibase 10.1103/PhysRevD.110.044039} {\bibfield  {journal} {\bibinfo
  {journal} {Phys. Rev. D}\ }\textbf {\bibinfo {volume} {110}},\ \bibinfo
  {pages} {044039} (\bibinfo {year} {2024})},\ \Eprint
  {http://arxiv.org/abs/2406.08363} {arXiv:2406.08363 [gr-qc]} \BibitemShut
  {NoStop}%
\bibitem [{\citenamefont {Albanesi}\ \emph
  {et~al.}(2025{\natexlab{a}})\citenamefont {Albanesi}, \citenamefont {Rashti},
  \citenamefont {Zappa}, \citenamefont {Gamba}, \citenamefont {Cook},
  \citenamefont {Daszuta}, \citenamefont {Bernuzzi}, \citenamefont {Nagar},\
  and\ \citenamefont {Radice}}]{Albanesi:2024xus}%
  \BibitemOpen
  \bibfield  {author} {\bibinfo {author} {\bibfnamefont {S.}~\bibnamefont
  {Albanesi}}, \bibinfo {author} {\bibfnamefont {A.}~\bibnamefont {Rashti}},
  \bibinfo {author} {\bibfnamefont {F.}~\bibnamefont {Zappa}}, \bibinfo
  {author} {\bibfnamefont {R.}~\bibnamefont {Gamba}}, \bibinfo {author}
  {\bibfnamefont {W.}~\bibnamefont {Cook}}, \bibinfo {author} {\bibfnamefont
  {B.}~\bibnamefont {Daszuta}}, \bibinfo {author} {\bibfnamefont
  {S.}~\bibnamefont {Bernuzzi}}, \bibinfo {author} {\bibfnamefont
  {A.}~\bibnamefont {Nagar}}, \ and\ \bibinfo {author} {\bibfnamefont
  {D.}~\bibnamefont {Radice}},\ }\href {\doibase 10.1103/PhysRevD.111.024069}
  {\bibfield  {journal} {\bibinfo  {journal} {Phys. Rev. D}\ }\textbf {\bibinfo
  {volume} {111}},\ \bibinfo {pages} {024069} (\bibinfo {year}
  {2025}{\natexlab{a}})},\ \Eprint {http://arxiv.org/abs/2405.20398}
  {arXiv:2405.20398 [gr-qc]} \BibitemShut {NoStop}%
\bibitem [{\citenamefont {Swain}\ \emph {et~al.}(2025)\citenamefont {Swain},
  \citenamefont {Pratten},\ and\ \citenamefont {Schmidt}}]{Swain:2024ngs}%
  \BibitemOpen
  \bibfield  {author} {\bibinfo {author} {\bibfnamefont {S.}~\bibnamefont
  {Swain}}, \bibinfo {author} {\bibfnamefont {G.}~\bibnamefont {Pratten}}, \
  and\ \bibinfo {author} {\bibfnamefont {P.}~\bibnamefont {Schmidt}},\ }\href
  {\doibase 10.1103/PhysRevD.111.064048} {\bibfield  {journal} {\bibinfo
  {journal} {Phys. Rev. D}\ }\textbf {\bibinfo {volume} {111}},\ \bibinfo
  {pages} {064048} (\bibinfo {year} {2025})},\ \Eprint
  {http://arxiv.org/abs/2411.09652} {arXiv:2411.09652 [gr-qc]} \BibitemShut
  {NoStop}%
\bibitem [{\citenamefont {Smith}\ \emph
  {et~al.}(2025{\natexlab{a}})\citenamefont {Smith}, \citenamefont
  {Paschalidis},\ and\ \citenamefont {Bozzola}}]{Smith:2024vka}%
  \BibitemOpen
  \bibfield  {author} {\bibinfo {author} {\bibfnamefont {M.~A.~M.}\
  \bibnamefont {Smith}}, \bibinfo {author} {\bibfnamefont {V.}~\bibnamefont
  {Paschalidis}}, \ and\ \bibinfo {author} {\bibfnamefont {G.}~\bibnamefont
  {Bozzola}},\ }\href {\doibase 10.1103/PhysRevD.111.104031} {\bibfield
  {journal} {\bibinfo  {journal} {Phys. Rev. D}\ }\textbf {\bibinfo {volume}
  {111}},\ \bibinfo {pages} {104031} (\bibinfo {year} {2025}{\natexlab{a}})},\
  \Eprint {http://arxiv.org/abs/2411.11960} {arXiv:2411.11960 [gr-qc]}
  \BibitemShut {NoStop}%
\bibitem [{\citenamefont {Smith}\ \emph
  {et~al.}(2025{\natexlab{b}})\citenamefont {Smith}, \citenamefont
  {Paschalidis},\ and\ \citenamefont {Bozzola}}]{Smith:2024nwq}%
  \BibitemOpen
  \bibfield  {author} {\bibinfo {author} {\bibfnamefont {M.~A.~M.}\
  \bibnamefont {Smith}}, \bibinfo {author} {\bibfnamefont {V.}~\bibnamefont
  {Paschalidis}}, \ and\ \bibinfo {author} {\bibfnamefont {G.}~\bibnamefont
  {Bozzola}},\ }\href {\doibase 10.1103/PhysRevD.111.104032} {\bibfield
  {journal} {\bibinfo  {journal} {Phys. Rev. D}\ }\textbf {\bibinfo {volume}
  {111}},\ \bibinfo {pages} {104032} (\bibinfo {year} {2025}{\natexlab{b}})},\
  \Eprint {http://arxiv.org/abs/2412.01881} {arXiv:2412.01881 [gr-qc]}
  \BibitemShut {NoStop}%
\bibitem [{\citenamefont {Fontbut\'e}\ \emph {et~al.}(2025)\citenamefont
  {Fontbut\'e}, \citenamefont {Andrade}, \citenamefont {Luna}, \citenamefont
  {Calder\'on~Bustillo}, \citenamefont {Morr\'as}, \citenamefont {Jaraba},
  \citenamefont {Garc\'\i{}a-Bellido},\ and\ \citenamefont
  {Izquierdo}}]{Fontbute:2024amb}%
  \BibitemOpen
  \bibfield  {author} {\bibinfo {author} {\bibfnamefont {J.}~\bibnamefont
  {Fontbut\'e}}, \bibinfo {author} {\bibfnamefont {T.}~\bibnamefont {Andrade}},
  \bibinfo {author} {\bibfnamefont {R.}~\bibnamefont {Luna}}, \bibinfo {author}
  {\bibfnamefont {J.}~\bibnamefont {Calder\'on~Bustillo}}, \bibinfo {author}
  {\bibfnamefont {G.}~\bibnamefont {Morr\'as}}, \bibinfo {author}
  {\bibfnamefont {S.}~\bibnamefont {Jaraba}}, \bibinfo {author} {\bibfnamefont
  {J.}~\bibnamefont {Garc\'\i{}a-Bellido}}, \ and\ \bibinfo {author}
  {\bibfnamefont {G.~L.}\ \bibnamefont {Izquierdo}},\ }\href {\doibase
  10.1103/PhysRevD.111.044024} {\bibfield  {journal} {\bibinfo  {journal}
  {Phys. Rev. D}\ }\textbf {\bibinfo {volume} {111}},\ \bibinfo {pages}
  {044024} (\bibinfo {year} {2025})},\ \Eprint
  {http://arxiv.org/abs/2409.16742} {arXiv:2409.16742 [gr-qc]} \BibitemShut
  {NoStop}%
\bibitem [{\citenamefont {Albanesi}\ \emph
  {et~al.}(2025{\natexlab{b}})\citenamefont {Albanesi}, \citenamefont {Gamba},
  \citenamefont {Bernuzzi}, \citenamefont {Fontbut\'e}, \citenamefont
  {Gonzalez},\ and\ \citenamefont {Nagar}}]{Albanesi:2025txj}%
  \BibitemOpen
  \bibfield  {author} {\bibinfo {author} {\bibfnamefont {S.}~\bibnamefont
  {Albanesi}}, \bibinfo {author} {\bibfnamefont {R.}~\bibnamefont {Gamba}},
  \bibinfo {author} {\bibfnamefont {S.}~\bibnamefont {Bernuzzi}}, \bibinfo
  {author} {\bibfnamefont {J.}~\bibnamefont {Fontbut\'e}}, \bibinfo {author}
  {\bibfnamefont {A.}~\bibnamefont {Gonzalez}}, \ and\ \bibinfo {author}
  {\bibfnamefont {A.}~\bibnamefont {Nagar}},\ }\href@noop {} {\  (\bibinfo
  {year} {2025}{\natexlab{b}})},\ \Eprint {http://arxiv.org/abs/2503.14580}
  {arXiv:2503.14580 [gr-qc]} \BibitemShut {NoStop}%
\bibitem [{\citenamefont {Punturo}\ \emph {et~al.}(2010)\citenamefont
  {Punturo}, \citenamefont {Abernathy}, \citenamefont {Acernese}, \citenamefont
  {Allen}, \citenamefont {Andersson} \emph {et~al.}}]{Punturo:2010zz}%
  \BibitemOpen
  \bibfield  {author} {\bibinfo {author} {\bibfnamefont {M.}~\bibnamefont
  {Punturo}}, \bibinfo {author} {\bibfnamefont {M.}~\bibnamefont {Abernathy}},
  \bibinfo {author} {\bibfnamefont {F.}~\bibnamefont {Acernese}}, \bibinfo
  {author} {\bibfnamefont {B.}~\bibnamefont {Allen}}, \bibinfo {author}
  {\bibfnamefont {N.}~\bibnamefont {Andersson}},  \emph {et~al.},\ }\href
  {\doibase 10.1088/0264-9381/27/19/194002} {\bibfield  {journal} {\bibinfo
  {journal} {Class.Quant.Grav.}\ }\textbf {\bibinfo {volume} {27}},\ \bibinfo
  {pages} {194002} (\bibinfo {year} {2010})}\BibitemShut {NoStop}%
\bibitem [{\citenamefont {Amaro-Seoane}\ \emph {et~al.}(2017)\citenamefont
  {Amaro-Seoane} \emph {et~al.}}]{LISA:2017pwj}%
  \BibitemOpen
  \bibfield  {author} {\bibinfo {author} {\bibfnamefont {P.}~\bibnamefont
  {Amaro-Seoane}} \emph {et~al.} (\bibinfo {collaboration} {LISA}),\
  }\href@noop {} {\  (\bibinfo {year} {2017})},\ \Eprint
  {http://arxiv.org/abs/1702.00786} {arXiv:1702.00786 [astro-ph.IM]}
  \BibitemShut {NoStop}%
\bibitem [{\citenamefont {Reitze}\ \emph {et~al.}(2019)\citenamefont {Reitze}
  \emph {et~al.}}]{Reitze:2019iox}%
  \BibitemOpen
  \bibfield  {author} {\bibinfo {author} {\bibfnamefont {D.}~\bibnamefont
  {Reitze}} \emph {et~al.},\ }\href@noop {} {\bibfield  {journal} {\bibinfo
  {journal} {Bull. Am. Astron. Soc.}\ }\textbf {\bibinfo {volume} {51}},\
  \bibinfo {pages} {035} (\bibinfo {year} {2019})},\ \Eprint
  {http://arxiv.org/abs/1907.04833} {arXiv:1907.04833 [astro-ph.IM]}
  \BibitemShut {NoStop}%
\bibitem [{\citenamefont {Afshordi}\ \emph {et~al.}(2023)\citenamefont
  {Afshordi} \emph {et~al.}}]{LISAConsortiumWaveformWorkingGroup:2023arg}%
  \BibitemOpen
  \bibfield  {author} {\bibinfo {author} {\bibfnamefont {N.}~\bibnamefont
  {Afshordi}} \emph {et~al.} (\bibinfo {collaboration} {LISA Consortium
  Waveform Working Group}),\ }\href@noop {} {\  (\bibinfo {year} {2023})},\
  \Eprint {http://arxiv.org/abs/2311.01300} {arXiv:2311.01300 [gr-qc]}
  \BibitemShut {NoStop}%
\bibitem [{\citenamefont {Abac}\ \emph {et~al.}(2025)\citenamefont {Abac} \emph
  {et~al.}}]{Abac:2025saz}%
  \BibitemOpen
  \bibfield  {author} {\bibinfo {author} {\bibfnamefont {A.}~\bibnamefont
  {Abac}} \emph {et~al.},\ }\href@noop {} {\  (\bibinfo {year} {2025})},\
  \Eprint {http://arxiv.org/abs/2503.12263} {arXiv:2503.12263 [gr-qc]}
  \BibitemShut {NoStop}%
\bibitem [{\citenamefont {Buonanno}\ and\ \citenamefont
  {Damour}(1999)}]{Buonanno:1998gg}%
  \BibitemOpen
  \bibfield  {author} {\bibinfo {author} {\bibfnamefont {A.}~\bibnamefont
  {Buonanno}}\ and\ \bibinfo {author} {\bibfnamefont {T.}~\bibnamefont
  {Damour}},\ }\href {\doibase 10.1103/PhysRevD.59.084006} {\bibfield
  {journal} {\bibinfo  {journal} {Phys. Rev.}\ }\textbf {\bibinfo {volume}
  {D59}},\ \bibinfo {pages} {084006} (\bibinfo {year} {1999})},\ \Eprint
  {http://arxiv.org/abs/gr-qc/9811091} {arXiv:gr-qc/9811091} \BibitemShut
  {NoStop}%
\bibitem [{\citenamefont {Buonanno}\ and\ \citenamefont
  {Damour}(2000)}]{Buonanno:2000ef}%
  \BibitemOpen
  \bibfield  {author} {\bibinfo {author} {\bibfnamefont {A.}~\bibnamefont
  {Buonanno}}\ and\ \bibinfo {author} {\bibfnamefont {T.}~\bibnamefont
  {Damour}},\ }\href {\doibase 10.1103/PhysRevD.62.064015} {\bibfield
  {journal} {\bibinfo  {journal} {Phys. Rev.}\ }\textbf {\bibinfo {volume}
  {D62}},\ \bibinfo {pages} {064015} (\bibinfo {year} {2000})},\ \Eprint
  {http://arxiv.org/abs/gr-qc/0001013} {arXiv:gr-qc/0001013} \BibitemShut
  {NoStop}%
\bibitem [{\citenamefont {Nagar}\ and\ \citenamefont
  {Rettegno}(2021)}]{Nagar:2021xnh}%
  \BibitemOpen
  \bibfield  {author} {\bibinfo {author} {\bibfnamefont {A.}~\bibnamefont
  {Nagar}}\ and\ \bibinfo {author} {\bibfnamefont {P.}~\bibnamefont
  {Rettegno}},\ }\href {\doibase 10.1103/PhysRevD.104.104004} {\bibfield
  {journal} {\bibinfo  {journal} {Phys. Rev. D}\ }\textbf {\bibinfo {volume}
  {104}},\ \bibinfo {pages} {104004} (\bibinfo {year} {2021})},\ \Eprint
  {http://arxiv.org/abs/2108.02043} {arXiv:2108.02043 [gr-qc]} \BibitemShut
  {NoStop}%
\bibitem [{\citenamefont {Nagar}\ \emph {et~al.}(2025)\citenamefont {Nagar},
  \citenamefont {Chiaramello}, \citenamefont {Gamba}, \citenamefont {Albanesi},
  \citenamefont {Bernuzzi}, \citenamefont {Fantini}, \citenamefont {Panzeri},\
  and\ \citenamefont {Rettegno}}]{Nagar:2024oyk}%
  \BibitemOpen
  \bibfield  {author} {\bibinfo {author} {\bibfnamefont {A.}~\bibnamefont
  {Nagar}}, \bibinfo {author} {\bibfnamefont {D.}~\bibnamefont {Chiaramello}},
  \bibinfo {author} {\bibfnamefont {R.}~\bibnamefont {Gamba}}, \bibinfo
  {author} {\bibfnamefont {S.}~\bibnamefont {Albanesi}}, \bibinfo {author}
  {\bibfnamefont {S.}~\bibnamefont {Bernuzzi}}, \bibinfo {author}
  {\bibfnamefont {V.}~\bibnamefont {Fantini}}, \bibinfo {author} {\bibfnamefont
  {M.}~\bibnamefont {Panzeri}}, \ and\ \bibinfo {author} {\bibfnamefont
  {P.}~\bibnamefont {Rettegno}},\ }\href {\doibase 10.1103/PhysRevD.111.064050}
  {\bibfield  {journal} {\bibinfo  {journal} {Phys. Rev. D}\ }\textbf {\bibinfo
  {volume} {111}},\ \bibinfo {pages} {064050} (\bibinfo {year} {2025})},\
  \Eprint {http://arxiv.org/abs/2407.04762} {arXiv:2407.04762 [gr-qc]}
  \BibitemShut {NoStop}%
\bibitem [{\citenamefont {Pompili}\ \emph {et~al.}(2023)\citenamefont {Pompili}
  \emph {et~al.}}]{Pompili:2023tna}%
  \BibitemOpen
  \bibfield  {author} {\bibinfo {author} {\bibfnamefont {L.}~\bibnamefont
  {Pompili}} \emph {et~al.},\ }\href {\doibase 10.1103/PhysRevD.108.124035}
  {\bibfield  {journal} {\bibinfo  {journal} {Phys. Rev. D}\ }\textbf {\bibinfo
  {volume} {108}},\ \bibinfo {pages} {124035} (\bibinfo {year} {2023})},\
  \Eprint {http://arxiv.org/abs/2303.18039} {arXiv:2303.18039 [gr-qc]}
  \BibitemShut {NoStop}%
\bibitem [{\citenamefont {Gamboa}\ \emph {et~al.}(2024)\citenamefont {Gamboa}
  \emph {et~al.}}]{Gamboa:2024hli}%
  \BibitemOpen
  \bibfield  {author} {\bibinfo {author} {\bibfnamefont {A.}~\bibnamefont
  {Gamboa}} \emph {et~al.},\ }\href@noop {} {\  (\bibinfo {year} {2024})},\
  \Eprint {http://arxiv.org/abs/2412.12823} {arXiv:2412.12823 [gr-qc]}
  \BibitemShut {NoStop}%
\bibitem [{\citenamefont {K\"alin}\ and\ \citenamefont
  {Porto}(2020)}]{Kalin:2020mvi}%
  \BibitemOpen
  \bibfield  {author} {\bibinfo {author} {\bibfnamefont {G.}~\bibnamefont
  {K\"alin}}\ and\ \bibinfo {author} {\bibfnamefont {R.~A.}\ \bibnamefont
  {Porto}},\ }\href {\doibase 10.1007/JHEP11(2020)106} {\bibfield  {journal}
  {\bibinfo  {journal} {JHEP}\ }\textbf {\bibinfo {volume} {11}},\ \bibinfo
  {pages} {106} (\bibinfo {year} {2020})},\ \Eprint
  {http://arxiv.org/abs/2006.01184} {arXiv:2006.01184 [hep-th]} \BibitemShut
  {NoStop}%
\bibitem [{\citenamefont {K\"alin}\ \emph
  {et~al.}(2020{\natexlab{a}})\citenamefont {K\"alin}, \citenamefont {Liu},\
  and\ \citenamefont {Porto}}]{Kalin:2020fhe}%
  \BibitemOpen
  \bibfield  {author} {\bibinfo {author} {\bibfnamefont {G.}~\bibnamefont
  {K\"alin}}, \bibinfo {author} {\bibfnamefont {Z.}~\bibnamefont {Liu}}, \ and\
  \bibinfo {author} {\bibfnamefont {R.~A.}\ \bibnamefont {Porto}},\ }\href
  {\doibase 10.1103/PhysRevLett.125.261103} {\bibfield  {journal} {\bibinfo
  {journal} {Phys. Rev. Lett.}\ }\textbf {\bibinfo {volume} {125}},\ \bibinfo
  {pages} {261103} (\bibinfo {year} {2020}{\natexlab{a}})},\ \Eprint
  {http://arxiv.org/abs/2007.04977} {arXiv:2007.04977 [hep-th]} \BibitemShut
  {NoStop}%
\bibitem [{\citenamefont {Dlapa}\ \emph {et~al.}(2022)\citenamefont {Dlapa},
  \citenamefont {K\"alin}, \citenamefont {Liu},\ and\ \citenamefont
  {Porto}}]{Dlapa:2021vgp}%
  \BibitemOpen
  \bibfield  {author} {\bibinfo {author} {\bibfnamefont {C.}~\bibnamefont
  {Dlapa}}, \bibinfo {author} {\bibfnamefont {G.}~\bibnamefont {K\"alin}},
  \bibinfo {author} {\bibfnamefont {Z.}~\bibnamefont {Liu}}, \ and\ \bibinfo
  {author} {\bibfnamefont {R.~A.}\ \bibnamefont {Porto}},\ }\href {\doibase
  10.1103/PhysRevLett.128.161104} {\bibfield  {journal} {\bibinfo  {journal}
  {Phys. Rev. Lett.}\ }\textbf {\bibinfo {volume} {128}},\ \bibinfo {pages}
  {161104} (\bibinfo {year} {2022})},\ \Eprint
  {http://arxiv.org/abs/2112.11296} {arXiv:2112.11296 [hep-th]} \BibitemShut
  {NoStop}%
\bibitem [{\citenamefont {Dlapa}\ \emph {et~al.}(2023)\citenamefont {Dlapa},
  \citenamefont {K\"alin}, \citenamefont {Liu}, \citenamefont {Neef},\ and\
  \citenamefont {Porto}}]{Dlapa:2022lmu}%
  \BibitemOpen
  \bibfield  {author} {\bibinfo {author} {\bibfnamefont {C.}~\bibnamefont
  {Dlapa}}, \bibinfo {author} {\bibfnamefont {G.}~\bibnamefont {K\"alin}},
  \bibinfo {author} {\bibfnamefont {Z.}~\bibnamefont {Liu}}, \bibinfo {author}
  {\bibfnamefont {J.}~\bibnamefont {Neef}}, \ and\ \bibinfo {author}
  {\bibfnamefont {R.~A.}\ \bibnamefont {Porto}},\ }\href {\doibase
  10.1103/PhysRevLett.130.101401} {\bibfield  {journal} {\bibinfo  {journal}
  {Phys. Rev. Lett.}\ }\textbf {\bibinfo {volume} {130}},\ \bibinfo {pages}
  {101401} (\bibinfo {year} {2023})},\ \Eprint
  {http://arxiv.org/abs/2210.05541} {arXiv:2210.05541 [hep-th]} \BibitemShut
  {NoStop}%
\bibitem [{\citenamefont {Vines}(2018)}]{Vines:2017hyw}%
  \BibitemOpen
  \bibfield  {author} {\bibinfo {author} {\bibfnamefont {J.}~\bibnamefont
  {Vines}},\ }\href {\doibase 10.1088/1361-6382/aaa3a8} {\bibfield  {journal}
  {\bibinfo  {journal} {Class. Quant. Grav.}\ }\textbf {\bibinfo {volume}
  {35}},\ \bibinfo {pages} {084002} (\bibinfo {year} {2018})},\ \Eprint
  {http://arxiv.org/abs/1709.06016} {arXiv:1709.06016 [gr-qc]} \BibitemShut
  {NoStop}%
\bibitem [{\citenamefont {Guevara}\ \emph {et~al.}(2019)\citenamefont
  {Guevara}, \citenamefont {Ochirov},\ and\ \citenamefont
  {Vines}}]{Guevara:2019fsj}%
  \BibitemOpen
  \bibfield  {author} {\bibinfo {author} {\bibfnamefont {A.}~\bibnamefont
  {Guevara}}, \bibinfo {author} {\bibfnamefont {A.}~\bibnamefont {Ochirov}}, \
  and\ \bibinfo {author} {\bibfnamefont {J.}~\bibnamefont {Vines}},\ }\href
  {\doibase 10.1103/PhysRevD.100.104024} {\bibfield  {journal} {\bibinfo
  {journal} {Phys. Rev. D}\ }\textbf {\bibinfo {volume} {100}},\ \bibinfo
  {pages} {104024} (\bibinfo {year} {2019})},\ \Eprint
  {http://arxiv.org/abs/1906.10071} {arXiv:1906.10071 [hep-th]} \BibitemShut
  {NoStop}%
\bibitem [{\citenamefont {Bern}\ \emph {et~al.}(2019)\citenamefont {Bern},
  \citenamefont {Cheung}, \citenamefont {Roiban}, \citenamefont {Shen},
  \citenamefont {Solon},\ and\ \citenamefont {Zeng}}]{Bern:2019nnu}%
  \BibitemOpen
  \bibfield  {author} {\bibinfo {author} {\bibfnamefont {Z.}~\bibnamefont
  {Bern}}, \bibinfo {author} {\bibfnamefont {C.}~\bibnamefont {Cheung}},
  \bibinfo {author} {\bibfnamefont {R.}~\bibnamefont {Roiban}}, \bibinfo
  {author} {\bibfnamefont {C.-H.}\ \bibnamefont {Shen}}, \bibinfo {author}
  {\bibfnamefont {M.~P.}\ \bibnamefont {Solon}}, \ and\ \bibinfo {author}
  {\bibfnamefont {M.}~\bibnamefont {Zeng}},\ }\href {\doibase
  10.1103/PhysRevLett.122.201603} {\bibfield  {journal} {\bibinfo  {journal}
  {Phys. Rev. Lett.}\ }\textbf {\bibinfo {volume} {122}},\ \bibinfo {pages}
  {201603} (\bibinfo {year} {2019})},\ \Eprint
  {http://arxiv.org/abs/1901.04424} {arXiv:1901.04424 [hep-th]} \BibitemShut
  {NoStop}%
\bibitem [{\citenamefont {Bern}\ \emph
  {et~al.}(2021{\natexlab{a}})\citenamefont {Bern}, \citenamefont
  {Parra-Martinez}, \citenamefont {Roiban}, \citenamefont {Ruf}, \citenamefont
  {Shen}, \citenamefont {Solon},\ and\ \citenamefont {Zeng}}]{Bern:2021dqo}%
  \BibitemOpen
  \bibfield  {author} {\bibinfo {author} {\bibfnamefont {Z.}~\bibnamefont
  {Bern}}, \bibinfo {author} {\bibfnamefont {J.}~\bibnamefont
  {Parra-Martinez}}, \bibinfo {author} {\bibfnamefont {R.}~\bibnamefont
  {Roiban}}, \bibinfo {author} {\bibfnamefont {M.~S.}\ \bibnamefont {Ruf}},
  \bibinfo {author} {\bibfnamefont {C.-H.}\ \bibnamefont {Shen}}, \bibinfo
  {author} {\bibfnamefont {M.~P.}\ \bibnamefont {Solon}}, \ and\ \bibinfo
  {author} {\bibfnamefont {M.}~\bibnamefont {Zeng}},\ }\href {\doibase
  10.1103/PhysRevLett.126.171601} {\bibfield  {journal} {\bibinfo  {journal}
  {Phys. Rev. Lett.}\ }\textbf {\bibinfo {volume} {126}},\ \bibinfo {pages}
  {171601} (\bibinfo {year} {2021}{\natexlab{a}})},\ \Eprint
  {http://arxiv.org/abs/2101.07254} {arXiv:2101.07254 [hep-th]} \BibitemShut
  {NoStop}%
\bibitem [{\citenamefont {Bern}\ \emph {et~al.}(2022)\citenamefont {Bern},
  \citenamefont {Parra-Martinez}, \citenamefont {Roiban}, \citenamefont {Ruf},
  \citenamefont {Shen}, \citenamefont {Solon},\ and\ \citenamefont
  {Zeng}}]{Bern:2021yeh}%
  \BibitemOpen
  \bibfield  {author} {\bibinfo {author} {\bibfnamefont {Z.}~\bibnamefont
  {Bern}}, \bibinfo {author} {\bibfnamefont {J.}~\bibnamefont
  {Parra-Martinez}}, \bibinfo {author} {\bibfnamefont {R.}~\bibnamefont
  {Roiban}}, \bibinfo {author} {\bibfnamefont {M.~S.}\ \bibnamefont {Ruf}},
  \bibinfo {author} {\bibfnamefont {C.-H.}\ \bibnamefont {Shen}}, \bibinfo
  {author} {\bibfnamefont {M.~P.}\ \bibnamefont {Solon}}, \ and\ \bibinfo
  {author} {\bibfnamefont {M.}~\bibnamefont {Zeng}},\ }\href {\doibase
  10.1103/PhysRevLett.128.161103} {\bibfield  {journal} {\bibinfo  {journal}
  {Phys. Rev. Lett.}\ }\textbf {\bibinfo {volume} {128}},\ \bibinfo {pages}
  {161103} (\bibinfo {year} {2022})},\ \Eprint
  {http://arxiv.org/abs/2112.10750} {arXiv:2112.10750 [hep-th]} \BibitemShut
  {NoStop}%
\bibitem [{\citenamefont {Manohar}\ \emph {et~al.}(2022)\citenamefont
  {Manohar}, \citenamefont {Ridgway},\ and\ \citenamefont
  {Shen}}]{Manohar:2022dea}%
  \BibitemOpen
  \bibfield  {author} {\bibinfo {author} {\bibfnamefont {A.~V.}\ \bibnamefont
  {Manohar}}, \bibinfo {author} {\bibfnamefont {A.~K.}\ \bibnamefont
  {Ridgway}}, \ and\ \bibinfo {author} {\bibfnamefont {C.-H.}\ \bibnamefont
  {Shen}},\ }\href {\doibase 10.1103/PhysRevLett.129.121601} {\bibfield
  {journal} {\bibinfo  {journal} {Phys. Rev. Lett.}\ }\textbf {\bibinfo
  {volume} {129}},\ \bibinfo {pages} {121601} (\bibinfo {year} {2022})},\
  \Eprint {http://arxiv.org/abs/2203.04283} {arXiv:2203.04283 [hep-th]}
  \BibitemShut {NoStop}%
\bibitem [{\citenamefont {Koemans~Collado}\ \emph {et~al.}(2019)\citenamefont
  {Koemans~Collado}, \citenamefont {Di~Vecchia},\ and\ \citenamefont
  {Russo}}]{KoemansCollado:2019ggb}%
  \BibitemOpen
  \bibfield  {author} {\bibinfo {author} {\bibfnamefont {A.}~\bibnamefont
  {Koemans~Collado}}, \bibinfo {author} {\bibfnamefont {P.}~\bibnamefont
  {Di~Vecchia}}, \ and\ \bibinfo {author} {\bibfnamefont {R.}~\bibnamefont
  {Russo}},\ }\href {\doibase 10.1103/PhysRevD.100.066028} {\bibfield
  {journal} {\bibinfo  {journal} {Phys. Rev. D}\ }\textbf {\bibinfo {volume}
  {100}},\ \bibinfo {pages} {066028} (\bibinfo {year} {2019})},\ \Eprint
  {http://arxiv.org/abs/1904.02667} {arXiv:1904.02667 [hep-th]} \BibitemShut
  {NoStop}%
\bibitem [{\citenamefont {Di~Vecchia}\ \emph {et~al.}(2020)\citenamefont
  {Di~Vecchia}, \citenamefont {Naculich}, \citenamefont {Russo}, \citenamefont
  {Veneziano},\ and\ \citenamefont {White}}]{DiVecchia:2019kta}%
  \BibitemOpen
  \bibfield  {author} {\bibinfo {author} {\bibfnamefont {P.}~\bibnamefont
  {Di~Vecchia}}, \bibinfo {author} {\bibfnamefont {S.~G.}\ \bibnamefont
  {Naculich}}, \bibinfo {author} {\bibfnamefont {R.}~\bibnamefont {Russo}},
  \bibinfo {author} {\bibfnamefont {G.}~\bibnamefont {Veneziano}}, \ and\
  \bibinfo {author} {\bibfnamefont {C.~D.}\ \bibnamefont {White}},\ }\href
  {\doibase 10.1007/JHEP03(2020)173} {\bibfield  {journal} {\bibinfo  {journal}
  {JHEP}\ }\textbf {\bibinfo {volume} {03}},\ \bibinfo {pages} {173} (\bibinfo
  {year} {2020})},\ \Eprint {http://arxiv.org/abs/1911.11716} {arXiv:1911.11716
  [hep-th]} \BibitemShut {NoStop}%
\bibitem [{\citenamefont {Di~Vecchia}\ \emph
  {et~al.}(2021{\natexlab{a}})\citenamefont {Di~Vecchia}, \citenamefont
  {Heissenberg}, \citenamefont {Russo},\ and\ \citenamefont
  {Veneziano}}]{DiVecchia:2021bdo}%
  \BibitemOpen
  \bibfield  {author} {\bibinfo {author} {\bibfnamefont {P.}~\bibnamefont
  {Di~Vecchia}}, \bibinfo {author} {\bibfnamefont {C.}~\bibnamefont
  {Heissenberg}}, \bibinfo {author} {\bibfnamefont {R.}~\bibnamefont {Russo}},
  \ and\ \bibinfo {author} {\bibfnamefont {G.}~\bibnamefont {Veneziano}},\
  }\href {\doibase 10.1007/JHEP07(2021)169} {\bibfield  {journal} {\bibinfo
  {journal} {JHEP}\ }\textbf {\bibinfo {volume} {07}},\ \bibinfo {pages} {169}
  (\bibinfo {year} {2021}{\natexlab{a}})},\ \Eprint
  {http://arxiv.org/abs/2104.03256} {arXiv:2104.03256 [hep-th]} \BibitemShut
  {NoStop}%
\bibitem [{\citenamefont {Di~Vecchia}\ \emph
  {et~al.}(2021{\natexlab{b}})\citenamefont {Di~Vecchia}, \citenamefont
  {Heissenberg}, \citenamefont {Russo},\ and\ \citenamefont
  {Veneziano}}]{DiVecchia:2021ndb}%
  \BibitemOpen
  \bibfield  {author} {\bibinfo {author} {\bibfnamefont {P.}~\bibnamefont
  {Di~Vecchia}}, \bibinfo {author} {\bibfnamefont {C.}~\bibnamefont
  {Heissenberg}}, \bibinfo {author} {\bibfnamefont {R.}~\bibnamefont {Russo}},
  \ and\ \bibinfo {author} {\bibfnamefont {G.}~\bibnamefont {Veneziano}},\
  }\href {\doibase 10.1016/j.physletb.2021.136379} {\bibfield  {journal}
  {\bibinfo  {journal} {Phys. Lett. B}\ }\textbf {\bibinfo {volume} {818}},\
  \bibinfo {pages} {136379} (\bibinfo {year} {2021}{\natexlab{b}})},\ \Eprint
  {http://arxiv.org/abs/2101.05772} {arXiv:2101.05772 [hep-th]} \BibitemShut
  {NoStop}%
\bibitem [{\citenamefont {Bini}\ and\ \citenamefont
  {Damour}(2022)}]{Bini:2022wrq}%
  \BibitemOpen
  \bibfield  {author} {\bibinfo {author} {\bibfnamefont {D.}~\bibnamefont
  {Bini}}\ and\ \bibinfo {author} {\bibfnamefont {T.}~\bibnamefont {Damour}},\
  }\href {\doibase 10.1103/PhysRevD.106.124049} {\bibfield  {journal} {\bibinfo
   {journal} {Phys. Rev. D}\ }\textbf {\bibinfo {volume} {106}},\ \bibinfo
  {pages} {124049} (\bibinfo {year} {2022})},\ \Eprint
  {http://arxiv.org/abs/2211.06340} {arXiv:2211.06340 [gr-qc]} \BibitemShut
  {NoStop}%
\bibitem [{\citenamefont {Bini}\ \emph {et~al.}(2023)\citenamefont {Bini},
  \citenamefont {Damour},\ and\ \citenamefont {Geralico}}]{Bini:2022enm}%
  \BibitemOpen
  \bibfield  {author} {\bibinfo {author} {\bibfnamefont {D.}~\bibnamefont
  {Bini}}, \bibinfo {author} {\bibfnamefont {T.}~\bibnamefont {Damour}}, \ and\
  \bibinfo {author} {\bibfnamefont {A.}~\bibnamefont {Geralico}},\ }\href
  {\doibase 10.1103/PhysRevD.107.024012} {\bibfield  {journal} {\bibinfo
  {journal} {Phys. Rev. D}\ }\textbf {\bibinfo {volume} {107}},\ \bibinfo
  {pages} {024012} (\bibinfo {year} {2023})},\ \Eprint
  {http://arxiv.org/abs/2210.07165} {arXiv:2210.07165 [gr-qc]} \BibitemShut
  {NoStop}%
\bibitem [{\citenamefont {Jakobsen}\ \emph
  {et~al.}(2023{\natexlab{a}})\citenamefont {Jakobsen}, \citenamefont {Mogull},
  \citenamefont {Plefka}, \citenamefont {Sauer},\ and\ \citenamefont
  {Xu}}]{Jakobsen:2023ndj}%
  \BibitemOpen
  \bibfield  {author} {\bibinfo {author} {\bibfnamefont {G.~U.}\ \bibnamefont
  {Jakobsen}}, \bibinfo {author} {\bibfnamefont {G.}~\bibnamefont {Mogull}},
  \bibinfo {author} {\bibfnamefont {J.}~\bibnamefont {Plefka}}, \bibinfo
  {author} {\bibfnamefont {B.}~\bibnamefont {Sauer}}, \ and\ \bibinfo {author}
  {\bibfnamefont {Y.}~\bibnamefont {Xu}},\ }\href {\doibase
  10.1103/PhysRevLett.131.151401} {\bibfield  {journal} {\bibinfo  {journal}
  {Phys. Rev. Lett.}\ }\textbf {\bibinfo {volume} {131}},\ \bibinfo {pages}
  {151401} (\bibinfo {year} {2023}{\natexlab{a}})},\ \Eprint
  {http://arxiv.org/abs/2306.01714} {arXiv:2306.01714 [hep-th]} \BibitemShut
  {NoStop}%
\bibitem [{\citenamefont {Driesse}\ \emph {et~al.}(2024)\citenamefont
  {Driesse}, \citenamefont {Jakobsen}, \citenamefont {Mogull}, \citenamefont
  {Plefka}, \citenamefont {Sauer},\ and\ \citenamefont
  {Usovitsch}}]{Driesse:2024xad}%
  \BibitemOpen
  \bibfield  {author} {\bibinfo {author} {\bibfnamefont {M.}~\bibnamefont
  {Driesse}}, \bibinfo {author} {\bibfnamefont {G.~U.}\ \bibnamefont
  {Jakobsen}}, \bibinfo {author} {\bibfnamefont {G.}~\bibnamefont {Mogull}},
  \bibinfo {author} {\bibfnamefont {J.}~\bibnamefont {Plefka}}, \bibinfo
  {author} {\bibfnamefont {B.}~\bibnamefont {Sauer}}, \ and\ \bibinfo {author}
  {\bibfnamefont {J.}~\bibnamefont {Usovitsch}},\ }\href {\doibase
  10.1103/PhysRevLett.132.241402} {\bibfield  {journal} {\bibinfo  {journal}
  {Phys. Rev. Lett.}\ }\textbf {\bibinfo {volume} {132}},\ \bibinfo {pages}
  {241402} (\bibinfo {year} {2024})},\ \Eprint
  {http://arxiv.org/abs/2403.07781} {arXiv:2403.07781 [hep-th]} \BibitemShut
  {NoStop}%
\bibitem [{\citenamefont {Bini}\ \emph {et~al.}(2020)\citenamefont {Bini},
  \citenamefont {Damour},\ and\ \citenamefont {Geralico}}]{Bini:2020flp}%
  \BibitemOpen
  \bibfield  {author} {\bibinfo {author} {\bibfnamefont {D.}~\bibnamefont
  {Bini}}, \bibinfo {author} {\bibfnamefont {T.}~\bibnamefont {Damour}}, \ and\
  \bibinfo {author} {\bibfnamefont {A.}~\bibnamefont {Geralico}},\ }\href
  {\doibase 10.1103/PhysRevD.101.044039} {\bibfield  {journal} {\bibinfo
  {journal} {Phys. Rev. D}\ }\textbf {\bibinfo {volume} {101}},\ \bibinfo
  {pages} {044039} (\bibinfo {year} {2020})},\ \Eprint
  {http://arxiv.org/abs/2001.00352} {arXiv:2001.00352 [gr-qc]} \BibitemShut
  {NoStop}%
\bibitem [{\citenamefont {Bern}\ \emph
  {et~al.}(2021{\natexlab{b}})\citenamefont {Bern}, \citenamefont
  {Parra-Martinez}, \citenamefont {Roiban}, \citenamefont {Sawyer},\ and\
  \citenamefont {Shen}}]{Bern:2020uwk}%
  \BibitemOpen
  \bibfield  {author} {\bibinfo {author} {\bibfnamefont {Z.}~\bibnamefont
  {Bern}}, \bibinfo {author} {\bibfnamefont {J.}~\bibnamefont
  {Parra-Martinez}}, \bibinfo {author} {\bibfnamefont {R.}~\bibnamefont
  {Roiban}}, \bibinfo {author} {\bibfnamefont {E.}~\bibnamefont {Sawyer}}, \
  and\ \bibinfo {author} {\bibfnamefont {C.-H.}\ \bibnamefont {Shen}},\ }\href
  {\doibase 10.1007/JHEP05(2021)188} {\bibfield  {journal} {\bibinfo  {journal}
  {JHEP}\ }\textbf {\bibinfo {volume} {05}},\ \bibinfo {pages} {188} (\bibinfo
  {year} {2021}{\natexlab{b}})},\ \Eprint {http://arxiv.org/abs/2010.08559}
  {arXiv:2010.08559 [hep-th]} \BibitemShut {NoStop}%
\bibitem [{\citenamefont {Cheung}\ and\ \citenamefont
  {Solon}(2020)}]{Cheung:2020sdj}%
  \BibitemOpen
  \bibfield  {author} {\bibinfo {author} {\bibfnamefont {C.}~\bibnamefont
  {Cheung}}\ and\ \bibinfo {author} {\bibfnamefont {M.~P.}\ \bibnamefont
  {Solon}},\ }\href {\doibase 10.1103/PhysRevLett.125.191601} {\bibfield
  {journal} {\bibinfo  {journal} {Phys. Rev. Lett.}\ }\textbf {\bibinfo
  {volume} {125}},\ \bibinfo {pages} {191601} (\bibinfo {year} {2020})},\
  \Eprint {http://arxiv.org/abs/2006.06665} {arXiv:2006.06665 [hep-th]}
  \BibitemShut {NoStop}%
\bibitem [{\citenamefont {K\"alin}\ \emph
  {et~al.}(2020{\natexlab{b}})\citenamefont {K\"alin}, \citenamefont {Liu},\
  and\ \citenamefont {Porto}}]{Kalin:2020lmz}%
  \BibitemOpen
  \bibfield  {author} {\bibinfo {author} {\bibfnamefont {G.}~\bibnamefont
  {K\"alin}}, \bibinfo {author} {\bibfnamefont {Z.}~\bibnamefont {Liu}}, \ and\
  \bibinfo {author} {\bibfnamefont {R.~A.}\ \bibnamefont {Porto}},\ }\href
  {\doibase 10.1103/PhysRevD.102.124025} {\bibfield  {journal} {\bibinfo
  {journal} {Phys. Rev. D}\ }\textbf {\bibinfo {volume} {102}},\ \bibinfo
  {pages} {124025} (\bibinfo {year} {2020}{\natexlab{b}})},\ \Eprint
  {http://arxiv.org/abs/2008.06047} {arXiv:2008.06047 [hep-th]} \BibitemShut
  {NoStop}%
\bibitem [{\citenamefont {Jakobsen}\ \emph {et~al.}(2024)\citenamefont
  {Jakobsen}, \citenamefont {Mogull}, \citenamefont {Plefka},\ and\
  \citenamefont {Sauer}}]{Jakobsen:2023pvx}%
  \BibitemOpen
  \bibfield  {author} {\bibinfo {author} {\bibfnamefont {G.~U.}\ \bibnamefont
  {Jakobsen}}, \bibinfo {author} {\bibfnamefont {G.}~\bibnamefont {Mogull}},
  \bibinfo {author} {\bibfnamefont {J.}~\bibnamefont {Plefka}}, \ and\ \bibinfo
  {author} {\bibfnamefont {B.}~\bibnamefont {Sauer}},\ }\href {\doibase
  10.1103/PhysRevD.109.L041504} {\bibfield  {journal} {\bibinfo  {journal}
  {Phys. Rev. D}\ }\textbf {\bibinfo {volume} {109}},\ \bibinfo {pages}
  {L041504} (\bibinfo {year} {2024})},\ \Eprint
  {http://arxiv.org/abs/2312.00719} {arXiv:2312.00719 [hep-th]} \BibitemShut
  {NoStop}%
\bibitem [{\citenamefont {Kovacs}\ and\ \citenamefont
  {Thorne}(1977)}]{Kovacs:1977uw}%
  \BibitemOpen
  \bibfield  {author} {\bibinfo {author} {\bibfnamefont {S.~J.}\ \bibnamefont
  {Kovacs}}\ and\ \bibinfo {author} {\bibfnamefont {K.~S.}\ \bibnamefont
  {Thorne}},\ }\href {\doibase 10.1086/155576} {\bibfield  {journal} {\bibinfo
  {journal} {Astrophys. J.}\ }\textbf {\bibinfo {volume} {217}},\ \bibinfo
  {pages} {252} (\bibinfo {year} {1977})}\BibitemShut {NoStop}%
\bibitem [{\citenamefont {Brandhuber}\ \emph {et~al.}(2023)\citenamefont
  {Brandhuber}, \citenamefont {Brown}, \citenamefont {Chen}, \citenamefont
  {De~Angelis}, \citenamefont {Gowdy},\ and\ \citenamefont
  {Travaglini}}]{Brandhuber:2023hhy}%
  \BibitemOpen
  \bibfield  {author} {\bibinfo {author} {\bibfnamefont {A.}~\bibnamefont
  {Brandhuber}}, \bibinfo {author} {\bibfnamefont {G.~R.}\ \bibnamefont
  {Brown}}, \bibinfo {author} {\bibfnamefont {G.}~\bibnamefont {Chen}},
  \bibinfo {author} {\bibfnamefont {S.}~\bibnamefont {De~Angelis}}, \bibinfo
  {author} {\bibfnamefont {J.}~\bibnamefont {Gowdy}}, \ and\ \bibinfo {author}
  {\bibfnamefont {G.}~\bibnamefont {Travaglini}},\ }\href {\doibase
  10.1007/JHEP06(2023)048} {\bibfield  {journal} {\bibinfo  {journal} {JHEP}\
  }\textbf {\bibinfo {volume} {06}},\ \bibinfo {pages} {048} (\bibinfo {year}
  {2023})},\ \Eprint {http://arxiv.org/abs/2303.06111} {arXiv:2303.06111
  [hep-th]} \BibitemShut {NoStop}%
\bibitem [{\citenamefont {Georgoudis}\ \emph {et~al.}(2023)\citenamefont
  {Georgoudis}, \citenamefont {Heissenberg},\ and\ \citenamefont
  {Vazquez-Holm}}]{Georgoudis:2023lgf}%
  \BibitemOpen
  \bibfield  {author} {\bibinfo {author} {\bibfnamefont {A.}~\bibnamefont
  {Georgoudis}}, \bibinfo {author} {\bibfnamefont {C.}~\bibnamefont
  {Heissenberg}}, \ and\ \bibinfo {author} {\bibfnamefont {I.}~\bibnamefont
  {Vazquez-Holm}},\ }\href {\doibase 10.1007/JHEP06(2023)126} {\bibfield
  {journal} {\bibinfo  {journal} {JHEP}\ }\textbf {\bibinfo {volume} {2023}},\
  \bibinfo {pages} {126} (\bibinfo {year} {2023})},\ \Eprint
  {http://arxiv.org/abs/2303.07006} {arXiv:2303.07006 [hep-th]} \BibitemShut
  {NoStop}%
\bibitem [{\citenamefont {Khalil}\ \emph {et~al.}(2022)\citenamefont {Khalil},
  \citenamefont {Buonanno}, \citenamefont {Steinhoff},\ and\ \citenamefont
  {Vines}}]{Khalil:2022ylj}%
  \BibitemOpen
  \bibfield  {author} {\bibinfo {author} {\bibfnamefont {M.}~\bibnamefont
  {Khalil}}, \bibinfo {author} {\bibfnamefont {A.}~\bibnamefont {Buonanno}},
  \bibinfo {author} {\bibfnamefont {J.}~\bibnamefont {Steinhoff}}, \ and\
  \bibinfo {author} {\bibfnamefont {J.}~\bibnamefont {Vines}},\ }\href
  {\doibase 10.1103/PhysRevD.106.024042} {\bibfield  {journal} {\bibinfo
  {journal} {Phys. Rev. D}\ }\textbf {\bibinfo {volume} {106}},\ \bibinfo
  {pages} {024042} (\bibinfo {year} {2022})},\ \Eprint
  {http://arxiv.org/abs/2204.05047} {arXiv:2204.05047 [gr-qc]} \BibitemShut
  {NoStop}%
\bibitem [{\citenamefont {Damour}\ and\ \citenamefont
  {Rettegno}(2023)}]{Damour:2022ybd}%
  \BibitemOpen
  \bibfield  {author} {\bibinfo {author} {\bibfnamefont {T.}~\bibnamefont
  {Damour}}\ and\ \bibinfo {author} {\bibfnamefont {P.}~\bibnamefont
  {Rettegno}},\ }\href {\doibase 10.1103/PhysRevD.107.064051} {\bibfield
  {journal} {\bibinfo  {journal} {Phys. Rev. D}\ }\textbf {\bibinfo {volume}
  {107}},\ \bibinfo {pages} {064051} (\bibinfo {year} {2023})},\ \Eprint
  {http://arxiv.org/abs/2211.01399} {arXiv:2211.01399 [gr-qc]} \BibitemShut
  {NoStop}%
\bibitem [{\citenamefont {Buonanno}\ \emph
  {et~al.}(2024{\natexlab{a}})\citenamefont {Buonanno}, \citenamefont {~},\
  and\ \citenamefont {Mogull}}]{Buonanno:2024vkx}%
  \BibitemOpen
  \bibfield  {author} {\bibinfo {author} {\bibfnamefont {A.}~\bibnamefont
  {Buonanno}}, \bibinfo {author} {\bibfnamefont {G.~U.}\ \bibnamefont {~}}, \
  and\ \bibinfo {author} {\bibfnamefont {G.}~\bibnamefont {Mogull}},\ }\href
  {\doibase 10.1103/PhysRevD.110.044038} {\bibfield  {journal} {\bibinfo
  {journal} {Phys. Rev. D}\ }\textbf {\bibinfo {volume} {110}},\ \bibinfo
  {pages} {044038} (\bibinfo {year} {2024}{\natexlab{a}})},\ \Eprint
  {http://arxiv.org/abs/2402.12342} {arXiv:2402.12342 [gr-qc]} \BibitemShut
  {NoStop}%
\bibitem [{\citenamefont {Damour}(2016)}]{Damour:2016gwp}%
  \BibitemOpen
  \bibfield  {author} {\bibinfo {author} {\bibfnamefont {T.}~\bibnamefont
  {Damour}},\ }\href {\doibase 10.1103/PhysRevD.94.104015} {\bibfield
  {journal} {\bibinfo  {journal} {Phys. Rev.}\ }\textbf {\bibinfo {volume}
  {D94}},\ \bibinfo {pages} {104015} (\bibinfo {year} {2016})},\ \Eprint
  {http://arxiv.org/abs/1609.00354} {arXiv:1609.00354 [gr-qc]} \BibitemShut
  {NoStop}%
\bibitem [{\citenamefont {Buonanno}\ \emph
  {et~al.}(2024{\natexlab{b}})\citenamefont {Buonanno}, \citenamefont {Mogull},
  \citenamefont {Patil},\ and\ \citenamefont {Pompili}}]{Buonanno:2024byg}%
  \BibitemOpen
  \bibfield  {author} {\bibinfo {author} {\bibfnamefont {A.}~\bibnamefont
  {Buonanno}}, \bibinfo {author} {\bibfnamefont {G.}~\bibnamefont {Mogull}},
  \bibinfo {author} {\bibfnamefont {R.}~\bibnamefont {Patil}}, \ and\ \bibinfo
  {author} {\bibfnamefont {L.}~\bibnamefont {Pompili}},\ }\href {\doibase
  10.1103/PhysRevLett.133.211402} {\bibfield  {journal} {\bibinfo  {journal}
  {Phys. Rev. Lett.}\ }\textbf {\bibinfo {volume} {133}},\ \bibinfo {pages}
  {211402} (\bibinfo {year} {2024}{\natexlab{b}})},\ \Eprint
  {http://arxiv.org/abs/2405.19181} {arXiv:2405.19181 [gr-qc]} \BibitemShut
  {NoStop}%
\bibitem [{\citenamefont {Damour}\ \emph {et~al.}(2025)\citenamefont {Damour},
  \citenamefont {Nagar}, \citenamefont {Placidi},\ and\ \citenamefont
  {Rettegno}}]{Damour:2025uka}%
  \BibitemOpen
  \bibfield  {author} {\bibinfo {author} {\bibfnamefont {T.}~\bibnamefont
  {Damour}}, \bibinfo {author} {\bibfnamefont {A.}~\bibnamefont {Nagar}},
  \bibinfo {author} {\bibfnamefont {A.}~\bibnamefont {Placidi}}, \ and\
  \bibinfo {author} {\bibfnamefont {P.}~\bibnamefont {Rettegno}},\ }\href@noop
  {} {\  (\bibinfo {year} {2025})},\ \Eprint {http://arxiv.org/abs/2503.05487}
  {arXiv:2503.05487 [gr-qc]} \BibitemShut {NoStop}%
\bibitem [{\citenamefont {Baumgarte}\ \emph {et~al.}(1998)\citenamefont
  {Baumgarte}, \citenamefont {Cook}, \citenamefont {Scheel}, \citenamefont
  {Shapiro},\ and\ \citenamefont {Teukolsky}}]{Baumgarte:1997eg}%
  \BibitemOpen
  \bibfield  {author} {\bibinfo {author} {\bibfnamefont {T.}~\bibnamefont
  {Baumgarte}}, \bibinfo {author} {\bibfnamefont {G.}~\bibnamefont {Cook}},
  \bibinfo {author} {\bibfnamefont {M.}~\bibnamefont {Scheel}}, \bibinfo
  {author} {\bibfnamefont {S.}~\bibnamefont {Shapiro}}, \ and\ \bibinfo
  {author} {\bibfnamefont {S.}~\bibnamefont {Teukolsky}},\ }\href {\doibase
  10.1103/PhysRevD.57.7299} {\bibfield  {journal} {\bibinfo  {journal}
  {Phys.Rev.}\ }\textbf {\bibinfo {volume} {D57}},\ \bibinfo {pages} {7299}
  (\bibinfo {year} {1998})},\ \Eprint {http://arxiv.org/abs/gr-qc/9709026}
  {arXiv:gr-qc/9709026 [gr-qc]} \BibitemShut {NoStop}%
\bibitem [{\citenamefont {Pfeiffer}\ and\ \citenamefont
  {York}(2003)}]{Pfeiffer:2002iy}%
  \BibitemOpen
  \bibfield  {author} {\bibinfo {author} {\bibfnamefont {H.~P.}\ \bibnamefont
  {Pfeiffer}}\ and\ \bibinfo {author} {\bibfnamefont {J.~W.}\ \bibnamefont
  {York}, \bibfnamefont {Jr.}},\ }\href {\doibase 10.1103/PhysRevD.67.044022}
  {\bibfield  {journal} {\bibinfo  {journal} {Phys. Rev.}\ }\textbf {\bibinfo
  {volume} {D67}},\ \bibinfo {pages} {044022} (\bibinfo {year} {2003})},\
  \Eprint {http://arxiv.org/abs/gr-qc/0207095} {arXiv:gr-qc/0207095 [gr-qc]}
  \BibitemShut {NoStop}%
\bibitem [{\citenamefont {Isenberg}(2008)}]{Isenberg:2007zg}%
  \BibitemOpen
  \bibfield  {author} {\bibinfo {author} {\bibfnamefont {J.~A.}\ \bibnamefont
  {Isenberg}},\ }\href {\doibase 10.1142/S0218271808011997} {\bibfield
  {journal} {\bibinfo  {journal} {Int. J. Mod. Phys.}\ }\textbf {\bibinfo
  {volume} {D17}},\ \bibinfo {pages} {265} (\bibinfo {year} {2008})},\ \Eprint
  {http://arxiv.org/abs/gr-qc/0702113} {arXiv:gr-qc/0702113} \BibitemShut
  {NoStop}%
\bibitem [{\citenamefont {Moldenhauer}\ \emph {et~al.}(2014)\citenamefont
  {Moldenhauer}, \citenamefont {Markakis}, \citenamefont {Johnson-McDaniel},
  \citenamefont {Tichy},\ and\ \citenamefont
  {Br{\"u}gmann}}]{Moldenhauer:2014yaa}%
  \BibitemOpen
  \bibfield  {author} {\bibinfo {author} {\bibfnamefont {N.}~\bibnamefont
  {Moldenhauer}}, \bibinfo {author} {\bibfnamefont {C.~M.}\ \bibnamefont
  {Markakis}}, \bibinfo {author} {\bibfnamefont {N.~K.}\ \bibnamefont
  {Johnson-McDaniel}}, \bibinfo {author} {\bibfnamefont {W.}~\bibnamefont
  {Tichy}}, \ and\ \bibinfo {author} {\bibfnamefont {B.}~\bibnamefont
  {Br{\"u}gmann}},\ }\href {\doibase 10.1103/PhysRevD.90.084043} {\bibfield
  {journal} {\bibinfo  {journal} {Phys. Rev.}\ }\textbf {\bibinfo {volume}
  {D90}},\ \bibinfo {pages} {084043} (\bibinfo {year} {2014})},\ \Eprint
  {http://arxiv.org/abs/1408.4136} {arXiv:1408.4136 [gr-qc]} \BibitemShut
  {NoStop}%
\bibitem [{\citenamefont {Tichy}(2011)}]{Tichy:2011gw}%
  \BibitemOpen
  \bibfield  {author} {\bibinfo {author} {\bibfnamefont {W.}~\bibnamefont
  {Tichy}},\ }\href {\doibase 10.1103/PhysRevD.84.024041} {\bibfield  {journal}
  {\bibinfo  {journal} {Phys.Rev.}\ }\textbf {\bibinfo {volume} {D84}},\
  \bibinfo {pages} {024041} (\bibinfo {year} {2011})},\ \Eprint
  {http://arxiv.org/abs/1107.1440} {arXiv:1107.1440 [gr-qc]} \BibitemShut
  {NoStop}%
\bibitem [{\citenamefont {Tichy}(2012)}]{Tichy:2012rp}%
  \BibitemOpen
  \bibfield  {author} {\bibinfo {author} {\bibfnamefont {W.}~\bibnamefont
  {Tichy}},\ }\href {\doibase 10.1103/PhysRevD.86.064024} {\bibfield  {journal}
  {\bibinfo  {journal} {Phys. Rev. D}\ }\textbf {\bibinfo {volume} {86}},\
  \bibinfo {pages} {064024} (\bibinfo {year} {2012})},\ \Eprint
  {http://arxiv.org/abs/1209.5336} {arXiv:1209.5336 [gr-qc]} \BibitemShut
  {NoStop}%
\bibitem [{\citenamefont {Dietrich}\ \emph {et~al.}(2015)\citenamefont
  {Dietrich}, \citenamefont {Moldenhauer}, \citenamefont {Johnson-McDaniel},
  \citenamefont {Bernuzzi}, \citenamefont {Markakis}, \citenamefont
  {Br{\"u}gmann},\ and\ \citenamefont {Tichy}}]{Dietrich:2015pxa}%
  \BibitemOpen
  \bibfield  {author} {\bibinfo {author} {\bibfnamefont {T.}~\bibnamefont
  {Dietrich}}, \bibinfo {author} {\bibfnamefont {N.}~\bibnamefont
  {Moldenhauer}}, \bibinfo {author} {\bibfnamefont {N.~K.}\ \bibnamefont
  {Johnson-McDaniel}}, \bibinfo {author} {\bibfnamefont {S.}~\bibnamefont
  {Bernuzzi}}, \bibinfo {author} {\bibfnamefont {C.~M.}\ \bibnamefont
  {Markakis}}, \bibinfo {author} {\bibfnamefont {B.}~\bibnamefont
  {Br{\"u}gmann}}, \ and\ \bibinfo {author} {\bibfnamefont {W.}~\bibnamefont
  {Tichy}},\ }\href {\doibase 10.1103/PhysRevD.92.124007} {\bibfield  {journal}
  {\bibinfo  {journal} {Phys. Rev.}\ }\textbf {\bibinfo {volume} {D92}},\
  \bibinfo {pages} {124007} (\bibinfo {year} {2015})},\ \Eprint
  {http://arxiv.org/abs/1507.07100} {arXiv:1507.07100 [gr-qc]} \BibitemShut
  {NoStop}%
\bibitem [{\citenamefont {Tichy}\ \emph {et~al.}(2019)\citenamefont {Tichy},
  \citenamefont {Rashti}, \citenamefont {Dietrich}, \citenamefont {Dudi},\ and\
  \citenamefont {Brügmann}}]{Tichy:2019ouu}%
  \BibitemOpen
  \bibfield  {author} {\bibinfo {author} {\bibfnamefont {W.}~\bibnamefont
  {Tichy}}, \bibinfo {author} {\bibfnamefont {A.}~\bibnamefont {Rashti}},
  \bibinfo {author} {\bibfnamefont {T.}~\bibnamefont {Dietrich}}, \bibinfo
  {author} {\bibfnamefont {R.}~\bibnamefont {Dudi}}, \ and\ \bibinfo {author}
  {\bibfnamefont {B.}~\bibnamefont {Brügmann}},\ }\href {\doibase
  10.1103/PhysRevD.100.124046} {\bibfield  {journal} {\bibinfo  {journal}
  {Phys. Rev.}\ }\textbf {\bibinfo {volume} {D100}},\ \bibinfo {pages} {124046}
  (\bibinfo {year} {2019})},\ \Eprint {http://arxiv.org/abs/1910.09690}
  {arXiv:1910.09690 [gr-qc]} \BibitemShut {NoStop}%
\bibitem [{\citenamefont {Read}\ \emph {et~al.}(2009)\citenamefont {Read},
  \citenamefont {Lackey}, \citenamefont {Owen},\ and\ \citenamefont
  {Friedman}}]{Read:2008iy}%
  \BibitemOpen
  \bibfield  {author} {\bibinfo {author} {\bibfnamefont {J.~S.}\ \bibnamefont
  {Read}}, \bibinfo {author} {\bibfnamefont {B.~D.}\ \bibnamefont {Lackey}},
  \bibinfo {author} {\bibfnamefont {B.~J.}\ \bibnamefont {Owen}}, \ and\
  \bibinfo {author} {\bibfnamefont {J.~L.}\ \bibnamefont {Friedman}},\ }\href
  {\doibase 10.1103/PhysRevD.79.124032} {\bibfield  {journal} {\bibinfo
  {journal} {Phys. Rev.}\ }\textbf {\bibinfo {volume} {D79}},\ \bibinfo {pages}
  {124032} (\bibinfo {year} {2009})},\ \Eprint {http://arxiv.org/abs/0812.2163}
  {arXiv:0812.2163 [astro-ph]} \BibitemShut {NoStop}%
\bibitem [{\citenamefont {Damour}\ and\ \citenamefont
  {Nagar}(2010)}]{Damour:2009wj}%
  \BibitemOpen
  \bibfield  {author} {\bibinfo {author} {\bibfnamefont {T.}~\bibnamefont
  {Damour}}\ and\ \bibinfo {author} {\bibfnamefont {A.}~\bibnamefont {Nagar}},\
  }\href {\doibase 10.1103/PhysRevD.81.084016} {\bibfield  {journal} {\bibinfo
  {journal} {Phys. Rev.}\ }\textbf {\bibinfo {volume} {D81}},\ \bibinfo {pages}
  {084016} (\bibinfo {year} {2010})},\ \Eprint {http://arxiv.org/abs/0911.5041}
  {arXiv:0911.5041 [gr-qc]} \BibitemShut {NoStop}%
\bibitem [{\citenamefont {Damour}\ \emph {et~al.}(2012)\citenamefont {Damour},
  \citenamefont {Nagar},\ and\ \citenamefont {Villain}}]{Damour:2012yf}%
  \BibitemOpen
  \bibfield  {author} {\bibinfo {author} {\bibfnamefont {T.}~\bibnamefont
  {Damour}}, \bibinfo {author} {\bibfnamefont {A.}~\bibnamefont {Nagar}}, \
  and\ \bibinfo {author} {\bibfnamefont {L.}~\bibnamefont {Villain}},\ }\href
  {\doibase 10.1103/PhysRevD.85.123007} {\bibfield  {journal} {\bibinfo
  {journal} {Phys.Rev.}\ }\textbf {\bibinfo {volume} {D85}},\ \bibinfo {pages}
  {123007} (\bibinfo {year} {2012})},\ \Eprint {http://arxiv.org/abs/1203.4352}
  {arXiv:1203.4352 [gr-qc]} \BibitemShut {NoStop}%
\bibitem [{\citenamefont {Bernuzzi}\ and\ \citenamefont
  {Hilditch}(2010)}]{Bernuzzi:2009ex}%
  \BibitemOpen
  \bibfield  {author} {\bibinfo {author} {\bibfnamefont {S.}~\bibnamefont
  {Bernuzzi}}\ and\ \bibinfo {author} {\bibfnamefont {D.}~\bibnamefont
  {Hilditch}},\ }\href {\doibase 10.1103/PhysRevD.81.084003} {\bibfield
  {journal} {\bibinfo  {journal} {Phys. Rev.}\ }\textbf {\bibinfo {volume}
  {D81}},\ \bibinfo {pages} {084003} (\bibinfo {year} {2010})},\ \Eprint
  {http://arxiv.org/abs/0912.2920} {arXiv:0912.2920 [gr-qc]} \BibitemShut
  {NoStop}%
\bibitem [{\citenamefont {Hilditch}\ \emph {et~al.}(2013)\citenamefont
  {Hilditch}, \citenamefont {Bernuzzi}, \citenamefont {Thierfelder},
  \citenamefont {Cao}, \citenamefont {Tichy},\ and\ \citenamefont
  {Bruegmann}}]{Hilditch:2012fp}%
  \BibitemOpen
  \bibfield  {author} {\bibinfo {author} {\bibfnamefont {D.}~\bibnamefont
  {Hilditch}}, \bibinfo {author} {\bibfnamefont {S.}~\bibnamefont {Bernuzzi}},
  \bibinfo {author} {\bibfnamefont {M.}~\bibnamefont {Thierfelder}}, \bibinfo
  {author} {\bibfnamefont {Z.}~\bibnamefont {Cao}}, \bibinfo {author}
  {\bibfnamefont {W.}~\bibnamefont {Tichy}}, \ and\ \bibinfo {author}
  {\bibfnamefont {B.}~\bibnamefont {Bruegmann}},\ }\href {\doibase
  10.1103/PhysRevD.88.084057} {\bibfield  {journal} {\bibinfo  {journal} {Phys.
  Rev.}\ }\textbf {\bibinfo {volume} {D88}},\ \bibinfo {pages} {084057}
  (\bibinfo {year} {2013})},\ \Eprint {http://arxiv.org/abs/1212.2901}
  {arXiv:1212.2901 [gr-qc]} \BibitemShut {NoStop}%
\bibitem [{\citenamefont {Alcubierre}\ \emph {et~al.}(2005)\citenamefont
  {Alcubierre}, \citenamefont {Br{\"u}gmann}, \citenamefont {Diener},
  \citenamefont {Guzman}, \citenamefont {Hawke} \emph
  {et~al.}}]{Alcubierre:2004hr}%
  \BibitemOpen
  \bibfield  {author} {\bibinfo {author} {\bibfnamefont {M.}~\bibnamefont
  {Alcubierre}}, \bibinfo {author} {\bibfnamefont {B.}~\bibnamefont
  {Br{\"u}gmann}}, \bibinfo {author} {\bibfnamefont {P.}~\bibnamefont
  {Diener}}, \bibinfo {author} {\bibfnamefont {F.~S.}\ \bibnamefont {Guzman}},
  \bibinfo {author} {\bibfnamefont {I.}~\bibnamefont {Hawke}},  \emph
  {et~al.},\ }\href {\doibase 10.1103/PhysRevD.72.044004} {\bibfield  {journal}
  {\bibinfo  {journal} {Phys.Rev.}\ }\textbf {\bibinfo {volume} {D72}},\
  \bibinfo {pages} {044004} (\bibinfo {year} {2005})},\ \Eprint
  {http://arxiv.org/abs/gr-qc/0411149} {arXiv:gr-qc/0411149 [gr-qc]}
  \BibitemShut {NoStop}%
\bibitem [{\citenamefont {Baker}\ \emph {et~al.}(2007)\citenamefont {Baker},
  \citenamefont {van Meter}, \citenamefont {McWilliams}, \citenamefont
  {Centrella},\ and\ \citenamefont {Kelly}}]{Baker:2006ha}%
  \BibitemOpen
  \bibfield  {author} {\bibinfo {author} {\bibfnamefont {J.~G.}\ \bibnamefont
  {Baker}}, \bibinfo {author} {\bibfnamefont {J.~R.}\ \bibnamefont {van
  Meter}}, \bibinfo {author} {\bibfnamefont {S.~T.}\ \bibnamefont
  {McWilliams}}, \bibinfo {author} {\bibfnamefont {J.}~\bibnamefont
  {Centrella}}, \ and\ \bibinfo {author} {\bibfnamefont {B.~J.}\ \bibnamefont
  {Kelly}},\ }\href {\doibase 10.1103/PhysRevLett.99.181101} {\bibfield
  {journal} {\bibinfo  {journal} {Phys.Rev.Lett.}\ }\textbf {\bibinfo {volume}
  {99}},\ \bibinfo {pages} {181101} (\bibinfo {year} {2007})},\ \Eprint
  {http://arxiv.org/abs/gr-qc/0612024} {arXiv:gr-qc/0612024 [gr-qc]}
  \BibitemShut {NoStop}%
\bibitem [{\citenamefont {Bowen}\ and\ \citenamefont
  {York}(1980)}]{Bowen:1980yu}%
  \BibitemOpen
  \bibfield  {author} {\bibinfo {author} {\bibfnamefont {J.~M.}\ \bibnamefont
  {Bowen}}\ and\ \bibinfo {author} {\bibfnamefont {J.~W.}\ \bibnamefont {York},
  \bibfnamefont {Jr.}},\ }\href {\doibase 10.1103/PhysRevD.21.2047} {\bibfield
  {journal} {\bibinfo  {journal} {Phys. Rev.}\ }\textbf {\bibinfo {volume}
  {D21}},\ \bibinfo {pages} {2047} (\bibinfo {year} {1980})}\BibitemShut
  {NoStop}%
\bibitem [{\citenamefont {Brandt}\ and\ \citenamefont
  {Br{\"u}gmann}(1997)}]{Brandt:1997tf}%
  \BibitemOpen
  \bibfield  {author} {\bibinfo {author} {\bibfnamefont {S.}~\bibnamefont
  {Brandt}}\ and\ \bibinfo {author} {\bibfnamefont {B.}~\bibnamefont
  {Br{\"u}gmann}},\ }\href {\doibase 10.1103/PhysRevLett.78.3606} {\bibfield
  {journal} {\bibinfo  {journal} {Phys. Rev. Lett.}\ }\textbf {\bibinfo
  {volume} {78}},\ \bibinfo {pages} {3606} (\bibinfo {year} {1997})},\ \Eprint
  {http://arxiv.org/abs/gr-qc/9703066} {arXiv:gr-qc/9703066} \BibitemShut
  {NoStop}%
\bibitem [{\citenamefont {Ansorg}\ \emph {et~al.}(2004)\citenamefont {Ansorg},
  \citenamefont {Br{\"u}gmann},\ and\ \citenamefont {Tichy}}]{Ansorg:2004ds}%
  \BibitemOpen
  \bibfield  {author} {\bibinfo {author} {\bibfnamefont {M.}~\bibnamefont
  {Ansorg}}, \bibinfo {author} {\bibfnamefont {B.}~\bibnamefont
  {Br{\"u}gmann}}, \ and\ \bibinfo {author} {\bibfnamefont {W.}~\bibnamefont
  {Tichy}},\ }\href {\doibase 10.1103/PhysRevD.70.064011} {\bibfield  {journal}
  {\bibinfo  {journal} {Phys. Rev.}\ }\textbf {\bibinfo {volume} {D70}},\
  \bibinfo {pages} {064011} (\bibinfo {year} {2004})},\ \Eprint
  {http://arxiv.org/abs/gr-qc/0404056} {arXiv:gr-qc/0404056} \BibitemShut
  {NoStop}%
\bibitem [{\citenamefont {Br{\"u}gmann}\ \emph {et~al.}(2008)\citenamefont
  {Br{\"u}gmann}, \citenamefont {Gonzalez}, \citenamefont {Hannam},
  \citenamefont {Husa}, \citenamefont {Sperhake} \emph
  {et~al.}}]{Brugmann:2008zz}%
  \BibitemOpen
  \bibfield  {author} {\bibinfo {author} {\bibfnamefont {B.}~\bibnamefont
  {Br{\"u}gmann}}, \bibinfo {author} {\bibfnamefont {J.~A.}\ \bibnamefont
  {Gonzalez}}, \bibinfo {author} {\bibfnamefont {M.}~\bibnamefont {Hannam}},
  \bibinfo {author} {\bibfnamefont {S.}~\bibnamefont {Husa}}, \bibinfo {author}
  {\bibfnamefont {U.}~\bibnamefont {Sperhake}},  \emph {et~al.},\ }\href
  {\doibase 10.1103/PhysRevD.77.024027} {\bibfield  {journal} {\bibinfo
  {journal} {Phys.Rev.}\ }\textbf {\bibinfo {volume} {D77}},\ \bibinfo {pages}
  {024027} (\bibinfo {year} {2008})},\ \Eprint
  {http://arxiv.org/abs/gr-qc/0610128} {arXiv:gr-qc/0610128 [gr-qc]}
  \BibitemShut {NoStop}%
\bibitem [{\citenamefont {Thierfelder}\ \emph {et~al.}(2011)\citenamefont
  {Thierfelder}, \citenamefont {Bernuzzi},\ and\ \citenamefont
  {Br{\"u}gmann}}]{Thierfelder:2011yi}%
  \BibitemOpen
  \bibfield  {author} {\bibinfo {author} {\bibfnamefont {M.}~\bibnamefont
  {Thierfelder}}, \bibinfo {author} {\bibfnamefont {S.}~\bibnamefont
  {Bernuzzi}}, \ and\ \bibinfo {author} {\bibfnamefont {B.}~\bibnamefont
  {Br{\"u}gmann}},\ }\href {\doibase 10.1103/PhysRevD.84.044012} {\bibfield
  {journal} {\bibinfo  {journal} {Phys.Rev.}\ }\textbf {\bibinfo {volume}
  {D84}},\ \bibinfo {pages} {044012} (\bibinfo {year} {2011})},\ \Eprint
  {http://arxiv.org/abs/1104.4751} {arXiv:1104.4751 [gr-qc]} \BibitemShut
  {NoStop}%
\bibitem [{\citenamefont {Bernuzzi}\ \emph {et~al.}(2012)\citenamefont
  {Bernuzzi}, \citenamefont {Nagar}, \citenamefont {Thierfelder},\ and\
  \citenamefont {Br{\"u}gmann}}]{Bernuzzi:2012ci}%
  \BibitemOpen
  \bibfield  {author} {\bibinfo {author} {\bibfnamefont {S.}~\bibnamefont
  {Bernuzzi}}, \bibinfo {author} {\bibfnamefont {A.}~\bibnamefont {Nagar}},
  \bibinfo {author} {\bibfnamefont {M.}~\bibnamefont {Thierfelder}}, \ and\
  \bibinfo {author} {\bibfnamefont {B.}~\bibnamefont {Br{\"u}gmann}},\ }\href
  {\doibase 10.1103/PhysRevD.86.044030} {\bibfield  {journal} {\bibinfo
  {journal} {Phys.Rev.}\ }\textbf {\bibinfo {volume} {D86}},\ \bibinfo {pages}
  {044030} (\bibinfo {year} {2012})},\ \Eprint {http://arxiv.org/abs/1205.3403}
  {arXiv:1205.3403 [gr-qc]} \BibitemShut {NoStop}%
\bibitem [{\citenamefont {Fontbut{\'e}}\ \emph {et~al.}(2025)\citenamefont
  {Fontbut{\'e}}, \citenamefont {Bernuzzi}, \citenamefont {Albanesi},
  \citenamefont {Radice}, \citenamefont {Rashti}, \citenamefont {Cook},
  \citenamefont {Daszuta},\ and\ \citenamefont {Nagar}}]{Fontbute:2025ixd}%
  \BibitemOpen
  \bibfield  {author} {\bibinfo {author} {\bibfnamefont {J.}~\bibnamefont
  {Fontbut{\'e}}}, \bibinfo {author} {\bibfnamefont {S.}~\bibnamefont
  {Bernuzzi}}, \bibinfo {author} {\bibfnamefont {S.}~\bibnamefont {Albanesi}},
  \bibinfo {author} {\bibfnamefont {D.}~\bibnamefont {Radice}}, \bibinfo
  {author} {\bibfnamefont {A.}~\bibnamefont {Rashti}}, \bibinfo {author}
  {\bibfnamefont {W.}~\bibnamefont {Cook}}, \bibinfo {author} {\bibfnamefont
  {B.}~\bibnamefont {Daszuta}}, \ and\ \bibinfo {author} {\bibfnamefont
  {A.}~\bibnamefont {Nagar}},\ }\href@noop {} {\  (\bibinfo {year} {2025})},\
  \Eprint {http://arxiv.org/abs/2508.03799} {arXiv:2508.03799 [gr-qc]}
  \BibitemShut {NoStop}%
\bibitem [{\citenamefont {Radice}\ \emph {et~al.}(2016)\citenamefont {Radice},
  \citenamefont {Galeazzi}, \citenamefont {Lippuner}, \citenamefont {Roberts},
  \citenamefont {Ott},\ and\ \citenamefont {Rezzolla}}]{Radice:2016dwd}%
  \BibitemOpen
  \bibfield  {author} {\bibinfo {author} {\bibfnamefont {D.}~\bibnamefont
  {Radice}}, \bibinfo {author} {\bibfnamefont {F.}~\bibnamefont {Galeazzi}},
  \bibinfo {author} {\bibfnamefont {J.}~\bibnamefont {Lippuner}}, \bibinfo
  {author} {\bibfnamefont {L.~F.}\ \bibnamefont {Roberts}}, \bibinfo {author}
  {\bibfnamefont {C.~D.}\ \bibnamefont {Ott}}, \ and\ \bibinfo {author}
  {\bibfnamefont {L.}~\bibnamefont {Rezzolla}},\ }\href {\doibase
  10.1093/mnras/stw1227} {\bibfield  {journal} {\bibinfo  {journal} {Mon. Not.
  Roy. Astron. Soc.}\ }\textbf {\bibinfo {volume} {460}},\ \bibinfo {pages}
  {3255} (\bibinfo {year} {2016})},\ \Eprint {http://arxiv.org/abs/1601.02426}
  {arXiv:1601.02426 [astro-ph.HE]} \BibitemShut {NoStop}%
\bibitem [{\citenamefont {Papenfort}\ \emph {et~al.}(2018)\citenamefont
  {Papenfort}, \citenamefont {Gold},\ and\ \citenamefont
  {Rezzolla}}]{Papenfort:2018bjk}%
  \BibitemOpen
  \bibfield  {author} {\bibinfo {author} {\bibfnamefont {L.~J.}\ \bibnamefont
  {Papenfort}}, \bibinfo {author} {\bibfnamefont {R.}~\bibnamefont {Gold}}, \
  and\ \bibinfo {author} {\bibfnamefont {L.}~\bibnamefont {Rezzolla}},\ }\href
  {\doibase 10.1103/PhysRevD.98.104028} {\bibfield  {journal} {\bibinfo
  {journal} {Phys. Rev. D}\ }\textbf {\bibinfo {volume} {98}},\ \bibinfo
  {pages} {104028} (\bibinfo {year} {2018})},\ \Eprint
  {http://arxiv.org/abs/1807.03795} {arXiv:1807.03795 [gr-qc]} \BibitemShut
  {NoStop}%
\bibitem [{\citenamefont {Chaurasia}\ \emph {et~al.}(2018)\citenamefont
  {Chaurasia}, \citenamefont {Dietrich}, \citenamefont {Johnson-McDaniel},
  \citenamefont {Ujevic}, \citenamefont {Tichy},\ and\ \citenamefont
  {Br\"ugmann}}]{Chaurasia:2018zhg}%
  \BibitemOpen
  \bibfield  {author} {\bibinfo {author} {\bibfnamefont {S.~V.}\ \bibnamefont
  {Chaurasia}}, \bibinfo {author} {\bibfnamefont {T.}~\bibnamefont {Dietrich}},
  \bibinfo {author} {\bibfnamefont {N.~K.}\ \bibnamefont {Johnson-McDaniel}},
  \bibinfo {author} {\bibfnamefont {M.}~\bibnamefont {Ujevic}}, \bibinfo
  {author} {\bibfnamefont {W.}~\bibnamefont {Tichy}}, \ and\ \bibinfo {author}
  {\bibfnamefont {B.}~\bibnamefont {Br\"ugmann}},\ }\href {\doibase
  10.1103/PhysRevD.98.104005} {\bibfield  {journal} {\bibinfo  {journal} {Phys.
  Rev. D}\ }\textbf {\bibinfo {volume} {98}},\ \bibinfo {pages} {104005}
  (\bibinfo {year} {2018})},\ \Eprint {http://arxiv.org/abs/1807.06857}
  {arXiv:1807.06857 [gr-qc]} \BibitemShut {NoStop}%
\bibitem [{\citenamefont {Neuweiler}\ \emph {et~al.}(2025)\citenamefont
  {Neuweiler}, \citenamefont {Dietrich},\ and\ \citenamefont
  {Br\"ugmann}}]{Neuweiler:2025lte}%
  \BibitemOpen
  \bibfield  {author} {\bibinfo {author} {\bibfnamefont {A.}~\bibnamefont
  {Neuweiler}}, \bibinfo {author} {\bibfnamefont {T.}~\bibnamefont {Dietrich}},
  \ and\ \bibinfo {author} {\bibfnamefont {B.}~\bibnamefont {Br\"ugmann}},\
  }\href@noop {} {\  (\bibinfo {year} {2025})},\ \Eprint
  {http://arxiv.org/abs/2504.10228} {arXiv:2504.10228 [gr-qc]} \BibitemShut
  {NoStop}%
\bibitem [{\citenamefont {Gamba}\ \emph {et~al.}(2023)\citenamefont {Gamba}
  \emph {et~al.}}]{Gamba:2023mww}%
  \BibitemOpen
  \bibfield  {author} {\bibinfo {author} {\bibfnamefont {R.}~\bibnamefont
  {Gamba}} \emph {et~al.},\ }\href@noop {} {\  (\bibinfo {year} {2023})},\
  \Eprint {http://arxiv.org/abs/2307.15125} {arXiv:2307.15125 [gr-qc]}
  \BibitemShut {NoStop}%
\bibitem [{\citenamefont {Bini}\ \emph {et~al.}(2012)\citenamefont {Bini},
  \citenamefont {Damour},\ and\ \citenamefont {Faye}}]{Bini:2012gu}%
  \BibitemOpen
  \bibfield  {author} {\bibinfo {author} {\bibfnamefont {D.}~\bibnamefont
  {Bini}}, \bibinfo {author} {\bibfnamefont {T.}~\bibnamefont {Damour}}, \ and\
  \bibinfo {author} {\bibfnamefont {G.}~\bibnamefont {Faye}},\ }\href {\doibase
  10.1103/PhysRevD.85.124034} {\bibfield  {journal} {\bibinfo  {journal}
  {Phys.Rev.}\ }\textbf {\bibinfo {volume} {D85}},\ \bibinfo {pages} {124034}
  (\bibinfo {year} {2012})},\ \Eprint {http://arxiv.org/abs/1202.3565}
  {arXiv:1202.3565 [gr-qc]} \BibitemShut {NoStop}%
\bibitem [{\citenamefont {Bernuzzi}\ \emph {et~al.}(2015)\citenamefont
  {Bernuzzi}, \citenamefont {Nagar}, \citenamefont {Dietrich},\ and\
  \citenamefont {Damour}}]{Bernuzzi:2014owa}%
  \BibitemOpen
  \bibfield  {author} {\bibinfo {author} {\bibfnamefont {S.}~\bibnamefont
  {Bernuzzi}}, \bibinfo {author} {\bibfnamefont {A.}~\bibnamefont {Nagar}},
  \bibinfo {author} {\bibfnamefont {T.}~\bibnamefont {Dietrich}}, \ and\
  \bibinfo {author} {\bibfnamefont {T.}~\bibnamefont {Damour}},\ }\href
  {\doibase 10.1103/PhysRevLett.114.161103} {\bibfield  {journal} {\bibinfo
  {journal} {Phys.Rev.Lett.}\ }\textbf {\bibinfo {volume} {114}},\ \bibinfo
  {pages} {161103} (\bibinfo {year} {2015})},\ \Eprint
  {http://arxiv.org/abs/1412.4553} {arXiv:1412.4553 [gr-qc]} \BibitemShut
  {NoStop}%
\bibitem [{\citenamefont {Akcay}\ \emph {et~al.}(2019)\citenamefont {Akcay},
  \citenamefont {Bernuzzi}, \citenamefont {Messina}, \citenamefont {Nagar},
  \citenamefont {Ortiz},\ and\ \citenamefont {Rettegno}}]{Akcay:2018yyh}%
  \BibitemOpen
  \bibfield  {author} {\bibinfo {author} {\bibfnamefont {S.}~\bibnamefont
  {Akcay}}, \bibinfo {author} {\bibfnamefont {S.}~\bibnamefont {Bernuzzi}},
  \bibinfo {author} {\bibfnamefont {F.}~\bibnamefont {Messina}}, \bibinfo
  {author} {\bibfnamefont {A.}~\bibnamefont {Nagar}}, \bibinfo {author}
  {\bibfnamefont {N.}~\bibnamefont {Ortiz}}, \ and\ \bibinfo {author}
  {\bibfnamefont {P.}~\bibnamefont {Rettegno}},\ }\href {\doibase
  10.1103/PhysRevD.99.044051} {\bibfield  {journal} {\bibinfo  {journal} {Phys.
  Rev.}\ }\textbf {\bibinfo {volume} {D99}},\ \bibinfo {pages} {044051}
  (\bibinfo {year} {2019})},\ \Eprint {http://arxiv.org/abs/1812.02744}
  {arXiv:1812.02744 [gr-qc]} \BibitemShut {NoStop}%
\bibitem [{\citenamefont {Jakobsen}\ \emph
  {et~al.}(2023{\natexlab{b}})\citenamefont {Jakobsen}, \citenamefont {Mogull},
  \citenamefont {Plefka},\ and\ \citenamefont {Sauer}}]{Jakobsen:2023hig}%
  \BibitemOpen
  \bibfield  {author} {\bibinfo {author} {\bibfnamefont {G.~U.}\ \bibnamefont
  {Jakobsen}}, \bibinfo {author} {\bibfnamefont {G.}~\bibnamefont {Mogull}},
  \bibinfo {author} {\bibfnamefont {J.}~\bibnamefont {Plefka}}, \ and\ \bibinfo
  {author} {\bibfnamefont {B.}~\bibnamefont {Sauer}},\ }\href {\doibase
  10.1103/PhysRevLett.131.241402} {\bibfield  {journal} {\bibinfo  {journal}
  {Phys. Rev. Lett.}\ }\textbf {\bibinfo {volume} {131}},\ \bibinfo {pages}
  {241402} (\bibinfo {year} {2023}{\natexlab{b}})},\ \Eprint
  {http://arxiv.org/abs/2308.11514} {arXiv:2308.11514 [hep-th]} \BibitemShut
  {NoStop}%
\bibitem [{\citenamefont {Campanelli}\ \emph
  {et~al.}(2006{\natexlab{b}})\citenamefont {Campanelli}, \citenamefont
  {Kelly},\ and\ \citenamefont {Lousto}}]{Campanelli:2005ia}%
  \BibitemOpen
  \bibfield  {author} {\bibinfo {author} {\bibfnamefont {M.}~\bibnamefont
  {Campanelli}}, \bibinfo {author} {\bibfnamefont {B.~J.}\ \bibnamefont
  {Kelly}}, \ and\ \bibinfo {author} {\bibfnamefont {C.~O.}\ \bibnamefont
  {Lousto}},\ }\href {\doibase 10.1103/PhysRevD.73.064005} {\bibfield
  {journal} {\bibinfo  {journal} {Phys. Rev. D}\ }\textbf {\bibinfo {volume}
  {73}},\ \bibinfo {pages} {064005} (\bibinfo {year} {2006}{\natexlab{b}})},\
  \Eprint {http://arxiv.org/abs/gr-qc/0510122} {arXiv:gr-qc/0510122}
  \BibitemShut {NoStop}%
\bibitem [{\citenamefont {Nakano}\ \emph {et~al.}(2015)\citenamefont {Nakano},
  \citenamefont {Healy}, \citenamefont {Lousto},\ and\ \citenamefont
  {Zlochower}}]{Nakano:2015pta}%
  \BibitemOpen
  \bibfield  {author} {\bibinfo {author} {\bibfnamefont {H.}~\bibnamefont
  {Nakano}}, \bibinfo {author} {\bibfnamefont {J.}~\bibnamefont {Healy}},
  \bibinfo {author} {\bibfnamefont {C.~O.}\ \bibnamefont {Lousto}}, \ and\
  \bibinfo {author} {\bibfnamefont {Y.}~\bibnamefont {Zlochower}},\ }\href
  {\doibase 10.1103/PhysRevD.91.104022} {\bibfield  {journal} {\bibinfo
  {journal} {Phys. Rev. D}\ }\textbf {\bibinfo {volume} {91}},\ \bibinfo
  {pages} {104022} (\bibinfo {year} {2015})},\ \Eprint
  {http://arxiv.org/abs/1503.00718} {arXiv:1503.00718 [gr-qc]} \BibitemShut
  {NoStop}%
\bibitem [{\citenamefont {Reisswig}\ and\ \citenamefont
  {Pollney}(2011)}]{Reisswig:2010di}%
  \BibitemOpen
  \bibfield  {author} {\bibinfo {author} {\bibfnamefont {C.}~\bibnamefont
  {Reisswig}}\ and\ \bibinfo {author} {\bibfnamefont {D.}~\bibnamefont
  {Pollney}},\ }\href {\doibase 10.1088/0264-9381/28/19/195015} {\bibfield
  {journal} {\bibinfo  {journal} {Class.Quant.Grav.}\ }\textbf {\bibinfo
  {volume} {28}},\ \bibinfo {pages} {195015} (\bibinfo {year} {2011})},\
  \Eprint {http://arxiv.org/abs/1006.1632} {arXiv:1006.1632 [gr-qc]}
  \BibitemShut {NoStop}%
\bibitem [{\citenamefont {Zel'dovich}\ and\ \citenamefont
  {Polnarev}(1974)}]{Zeldovich:1974gvh}%
  \BibitemOpen
  \bibfield  {author} {\bibinfo {author} {\bibfnamefont {Y.~B.}\ \bibnamefont
  {Zel'dovich}}\ and\ \bibinfo {author} {\bibfnamefont {A.~G.}\ \bibnamefont
  {Polnarev}},\ }\href@noop {} {\bibfield  {journal} {\bibinfo  {journal} {Sov.
  Astron.}\ }\textbf {\bibinfo {volume} {18}},\ \bibinfo {pages} {17} (\bibinfo
  {year} {1974})}\BibitemShut {NoStop}%
\bibitem [{\citenamefont {Braginsky}\ and\ \citenamefont
  {Grishchuk}(1985)}]{Braginsky:1985vlg}%
  \BibitemOpen
  \bibfield  {author} {\bibinfo {author} {\bibfnamefont {V.~B.}\ \bibnamefont
  {Braginsky}}\ and\ \bibinfo {author} {\bibfnamefont {L.~P.}\ \bibnamefont
  {Grishchuk}},\ }\href@noop {} {\bibfield  {journal} {\bibinfo  {journal}
  {Sov. Phys. JETP}\ }\textbf {\bibinfo {volume} {62}},\ \bibinfo {pages} {427}
  (\bibinfo {year} {1985})}\BibitemShut {NoStop}%
\bibitem [{\citenamefont {Chiaramello}\ and\ \citenamefont
  {Nagar}(2020)}]{Chiaramello:2020ehz}%
  \BibitemOpen
  \bibfield  {author} {\bibinfo {author} {\bibfnamefont {D.}~\bibnamefont
  {Chiaramello}}\ and\ \bibinfo {author} {\bibfnamefont {A.}~\bibnamefont
  {Nagar}},\ }\href {\doibase 10.1103/PhysRevD.101.101501} {\bibfield
  {journal} {\bibinfo  {journal} {Phys. Rev. D}\ }\textbf {\bibinfo {volume}
  {101}},\ \bibinfo {pages} {101501} (\bibinfo {year} {2020})},\ \Eprint
  {http://arxiv.org/abs/2001.11736} {arXiv:2001.11736 [gr-qc]} \BibitemShut
  {NoStop}%
\bibitem [{\citenamefont {Gold}\ \emph {et~al.}(2012)\citenamefont {Gold},
  \citenamefont {Bernuzzi}, \citenamefont {Thierfelder}, \citenamefont
  {Br{\"u}gmann},\ and\ \citenamefont {Pretorius}}]{Gold:2011df}%
  \BibitemOpen
  \bibfield  {author} {\bibinfo {author} {\bibfnamefont {R.}~\bibnamefont
  {Gold}}, \bibinfo {author} {\bibfnamefont {S.}~\bibnamefont {Bernuzzi}},
  \bibinfo {author} {\bibfnamefont {M.}~\bibnamefont {Thierfelder}}, \bibinfo
  {author} {\bibfnamefont {B.}~\bibnamefont {Br{\"u}gmann}}, \ and\ \bibinfo
  {author} {\bibfnamefont {F.}~\bibnamefont {Pretorius}},\ }\href {\doibase
  10.1103/PhysRevD.86.121501} {\bibfield  {journal} {\bibinfo  {journal}
  {Phys.Rev.}\ }\textbf {\bibinfo {volume} {D86}},\ \bibinfo {pages} {121501}
  (\bibinfo {year} {2012})},\ \Eprint {http://arxiv.org/abs/1109.5128}
  {arXiv:1109.5128 [gr-qc]} \BibitemShut {NoStop}%
\bibitem [{\citenamefont {Abbott}\ \emph {et~al.}(2017)\citenamefont {Abbott}
  \emph {et~al.}}]{TheLIGOScientific:2017qsa}%
  \BibitemOpen
  \bibfield  {author} {\bibinfo {author} {\bibfnamefont {B.~P.}\ \bibnamefont
  {Abbott}} \emph {et~al.} (\bibinfo {collaboration} {Virgo, LIGO
  Scientific}),\ }\href {\doibase 10.1103/PhysRevLett.119.161101} {\bibfield
  {journal} {\bibinfo  {journal} {Phys. Rev. Lett.}\ }\textbf {\bibinfo
  {volume} {119}},\ \bibinfo {pages} {161101} (\bibinfo {year} {2017})},\
  \Eprint {http://arxiv.org/abs/1710.05832} {arXiv:1710.05832 [gr-qc]}
  \BibitemShut {NoStop}%
\bibitem [{\citenamefont {De}\ \emph {et~al.}(2018)\citenamefont {De},
  \citenamefont {Finstad}, \citenamefont {Lattimer}, \citenamefont {Brown},
  \citenamefont {Berger},\ and\ \citenamefont {Biwer}}]{De:2018uhw}%
  \BibitemOpen
  \bibfield  {author} {\bibinfo {author} {\bibfnamefont {S.}~\bibnamefont
  {De}}, \bibinfo {author} {\bibfnamefont {D.}~\bibnamefont {Finstad}},
  \bibinfo {author} {\bibfnamefont {J.~M.}\ \bibnamefont {Lattimer}}, \bibinfo
  {author} {\bibfnamefont {D.~A.}\ \bibnamefont {Brown}}, \bibinfo {author}
  {\bibfnamefont {E.}~\bibnamefont {Berger}}, \ and\ \bibinfo {author}
  {\bibfnamefont {C.~M.}\ \bibnamefont {Biwer}},\ }\href {\doibase
  10.1103/PhysRevLett.121.259902, 10.1103/PhysRevLett.121.091102} {\bibfield
  {journal} {\bibinfo  {journal} {Phys. Rev. Lett.}\ }\textbf {\bibinfo
  {volume} {121}},\ \bibinfo {pages} {091102} (\bibinfo {year} {2018})},\
  \bibinfo {note} {[Erratum: Phys. Rev. Lett.121,no.25,259902(2018)]},\ \Eprint
  {http://arxiv.org/abs/1804.08583} {arXiv:1804.08583 [astro-ph.HE]}
  \BibitemShut {NoStop}%
\bibitem [{\citenamefont {Abbott}\ \emph {et~al.}(2019)\citenamefont {Abbott}
  \emph {et~al.}}]{LIGOScientific:2018hze}%
  \BibitemOpen
  \bibfield  {author} {\bibinfo {author} {\bibfnamefont {B.~P.}\ \bibnamefont
  {Abbott}} \emph {et~al.} (\bibinfo {collaboration} {LIGO Scientific,
  Virgo}),\ }\href {\doibase 10.1103/PhysRevX.9.011001} {\bibfield  {journal}
  {\bibinfo  {journal} {Phys. Rev. X}\ }\textbf {\bibinfo {volume} {9}},\
  \bibinfo {pages} {011001} (\bibinfo {year} {2019})},\ \Eprint
  {http://arxiv.org/abs/1805.11579} {arXiv:1805.11579 [gr-qc]} \BibitemShut
  {NoStop}%
\bibitem [{\citenamefont {Rosswog}\ \emph {et~al.}(2013)\citenamefont
  {Rosswog}, \citenamefont {Piran},\ and\ \citenamefont
  {Nakar}}]{Rosswog:2012wb}%
  \BibitemOpen
  \bibfield  {author} {\bibinfo {author} {\bibfnamefont {S.}~\bibnamefont
  {Rosswog}}, \bibinfo {author} {\bibfnamefont {T.}~\bibnamefont {Piran}}, \
  and\ \bibinfo {author} {\bibfnamefont {E.}~\bibnamefont {Nakar}},\ }\href
  {\doibase 10.1093/mnras/sts708} {\bibfield  {journal} {\bibinfo  {journal}
  {Mon. Not. Roy. Astron. Soc.}\ }\textbf {\bibinfo {volume} {430}},\ \bibinfo
  {pages} {2585} (\bibinfo {year} {2013})},\ \Eprint
  {http://arxiv.org/abs/1204.6240} {arXiv:1204.6240 [astro-ph.HE]} \BibitemShut
  {NoStop}%
\end{thebibliography}
\end{document}